\documentclass[showpacs,preprintnumbers,aps,prl,letterpaper,superscriptaddress,nofootinbib,tightenlines,floats,floatfix]{revtex4-2}
\usepackage{graphicx}
\usepackage{bm}
\usepackage{latexsym}
\usepackage{amsmath,amssymb}
\usepackage{amsfonts}
\usepackage[draft=false]{hyperref}
\usepackage[utf8]{inputenc}
\usepackage{amsthm}
\usepackage{mathrsfs}
\usepackage{ifpdf}
\usepackage{color}
\usepackage{epstopdf}
\usepackage[skip=2pt,font=scriptsize]{caption}
\usepackage{natbib}
\usepackage{blindtext}
\DeclareMathOperator{\sgn}{sgn}
\allowdisplaybreaks

\begin{document}
\title{Testing general covariance in effective models motivated by Loop Quantum Gravity}

\author{Juan Carlos Del \'Aguila\footnote{Corresponding author: jcdelaguila@xanum.uam.mx}} 
\email{jcdelaguila@xanum.uam.mx} 
\affiliation{Departamento de F\'isica, Universidad Aut\'onoma Metropolitana Iztapalapa, San Rafael Atlixco 186, CP 09340, Ciudad de M\'exico, M\'exico.}
\author{Hugo A. Morales-Técotl} 
\email{hugo@xanum.uam.mx} 
\affiliation{Departamento de F\'isica, Universidad Aut\'onoma Metropolitana Iztapalapa, San Rafael Atlixco 186, CP 09340, Ciudad de M\'exico, M\'exico.} 

\begin{abstract}
In this work we introduce a criterion for testing general covariance in effective quantum gravity theories. It adapts the analysis of invariance under general spacetime diffeomorphisms of the Einstein-Hilbert action to the case of effective canonical models. While the main purpose is to test models obtained in Loop Quantum Gravity, the criterion is not limited to those physical systems and may be applied to any canonically formulated modified theory of gravity. The approach here is hence not that of finding an effective model, but rather to examine a given one represented by a quantum corrected Hamiltonian. Specifically, we will apply the criterion to spherically symmetric spacetimes in vacuum with inverse triad and holonomy modifications that arise as a consequence of the loop quantization procedure. It is found that, in addition to the initial modifications of the Hamiltonian, quantum corrections of the classical metric itself are needed as well in order to obtain generally covariant models. A comparison with recent alternative criteria is included in the discussion. 
\end{abstract}


\date{Received: date / Accepted: date}

\maketitle

\date{\today}


\maketitle

\section{I. Introduction}

Loop Quantum Gravity (LQG) is one of the most prominent proposals that attempt to quantize gravity. The starting point of the theory is the classical description of General Relativity (GR) as a Hamiltonian formulation. At the quantum level, LQG is characterized by the replacement of the canonical algebra between conjugated phase space variables, with the so-called holonomy-flux algebra. The absence of the canonical algebra is due to the non-existence of the infinitesimal generator of translations, reflecting thus the discreteness of the spacetime geometry. See \cite{AshtekarLewandowski} for a complete review on the subject.

Over the past years, much progress has been accomplished within the field of LQG by considering symmetry reduced models that describe relevant physical situations \cite{Symm}. In this process, the complete classical theory is reduced to a certain symmetrical sector of interest and then quantized, as opposed to performing the symmetry reduction in an already quantized theory. For instance, in Loop Quantum Cosmology (LQC) one of the main interests is the analysis of spatially homogeneous spacetimes, be them isotropic \cite{LQC, Improved} or anisotropic \cite{SchwarzschildSingularity, Modesto, Revisited} (just to name a few seminal references). The standard procedure involves working from the onset in the mini-superspace, where only a finite number of gravitational degrees of freedom remain. The system therefore can be treated as a mechanical model, instead of a more complex field theory. Quantization is then limited only to those degrees of freedom within the mini-superspace. 

Effective models in LQG emerge as an attempt to incorporate physical characteristics that arise during the loop quantization procedure into a classical framework. In this case, they consist of a modified theory of gravity in the classical phase space. Examples of techniques that have been used to obtain this type of effective models include the computation of the expectation value of the Hamiltonian operator using Gaussian states, and the path integral approach \cite{PathIntegralSLQC, Hugo1}. The advantage of such models is that one can readily use the well-known mathematical tools of GR to study quantum effects in a semi-classical approximation. Some interesting results yielded by effective analysis in LQC are the boundedness of scalar quantities relevant to the appearance of singularities such as curvature invariants, expansion, and shear in homogeneous cosmologies \cite{Corichi, Joe, Hugo2}, and the resolution of those singularities as a big bounce \cite{Improved, Vandersloot, RobustnessBounce}. 

Careful attention must be taken, though, in order for the obtained effective model to be considered as an appropriate theory in classical phase space. Namely, it has to satisfy the inherent properties and principles of the classical gravity framework. In this context, general covariance stands as a crucial characteristic of the theory that an effective model should possess. In the mini-superspace cases of LQC, since all but a finite quantity of degrees of freedom are frozen, the issue of covariance becomes trivial. Nevertheless, when considering more complicated spacetimes like those which are spatially inhomogeneous, this issue becomes relevant. An example of the former is the loss of covariance in spherically symmetric effective models \cite{Deformed1, Deformed2}. The treatment of these spacetimes differs from the homogeneous ones, even from the start in their classical description, mainly due to the presence of field theory aspects of the model. In fact, they are sometimes referred to as midi-superspace models \cite{midi}. This constitutes an important matter since these systems can be used to describe the exterior region of black holes, as well as the inner most region of a charged black hole. Interesting physical questions can then be formulated such as the possible singularity resolution in those spacetimes and a semi-classical description of radiation. If one wishes to give an answer to these situations from the effective theory perspective, then the general covariance problem needs to be addressed. Some proposals of covariant effective models with this type of symmetry can be found in \cite{Gambini, asier, zhang, zhang2, brahma, belfaqih, Asier2}. In particular, \cite{zhang, brahma, belfaqih, Asier2} provide a covariance criterion based on the transformations generated by the sum of constraints of the Hamiltonian applied to the metric coefficients. If these gauge transformations coincide on-shell with those induced by diffeomorphisms, then the modified action is covariant with respect to the analyzed metric. In the context of effective theories, it has been seen that said coefficients, in terms of the canonical variables of the formalism, do not necessarily coincide with those of the classical theory. Because of this, the new covariant line element occasionally has been called an ``emergent'' line element \cite{belfaqih}. This idea will be continuously used throughout this paper. Additional references that deal with this problem will be mentioned, along with further necessary context, in the main body of this work.

Our aim in this paper is to provide an explicit test of general covariance in effective theories of quantum gravity. After that, the main focus will turn to the spherically symmetric model in LQG. The work is organized as follows. Section II serves as a brief review of general covariance in both the Lagrangian and Hamiltonian formulations of GR. In section III we develop a criterion to test covariance in effective models. It will be majorly based on the Lagrangian approach, although properly adapted to be applied to canonical theories, which are our principal interest. Next, section IV contains the main results of the paper. Here, we test general covariance in effective spherical symmetry with two distinct types of corrections that typically arise in LQG: inverse triad and holonomy modifications. In section V, a link between a covariant spherically symmetric effective model (a midi-superspace model) and an adapted Kantowski-Sachs model (a mini-superspace model) is established. Finally, conclusions are drawn in the last section of the work, as well as possible future directions to extend it. An appendix is included at the end of the paper which contains additional calculations related to section IV.

\section{II. General Covariance in the Classical Theory}

One of the most important features of GR, the classical theory of gravitation, is general covariance. This property refers to the principle that the laws of physics should not change under general coordinate transformations. In GR, general covariance is expressed in mathematical terms by the invariance under diffeomorphisms, i.e., differentiable, one-to-one and onto maps between manifolds $\phi:M\rightarrow M$, of its action functional. This property can be easily seen by using the Lagrangian formulation of the theory \cite{wald}. In the vacuum, which will be the case of interest in this paper, this formulation provides us with the Einstein-Hilbert action, 

\begin{equation}
S_G[g_{\mu\nu}]=\frac{1}{8\pi}\int_M\sqrt{-g}Rd^4x,
\label{EHS}
\end{equation}
where $g_{\mu\nu}$ is the spacetime metric, $g=\det[g_{\mu\nu}]$, and $R$ is the Ricci scalar. Variation of this action with respect to $g_{\mu\nu}$ yields, 

\begin{equation}
\delta S_G=\frac{1}{8\pi}\int_M\frac{\delta S_G}{\delta g_{\mu\nu}}\delta g_{\mu\nu}d^4x.
\end{equation}
Under a one-parameter family of diffeomorphisms $\phi_s$, such that $\xi^\mu$ is the vector field that generates $\phi_s$, we have that $\delta g_{\mu\nu}=\pounds_\xi g_{\mu\nu}$. Here, we are denoting the Lie derivative along the vector field $\xi^\mu$ as $\pounds_\xi$. Since $$\frac{\delta S_G}{\delta g_{\mu\nu}}=\sqrt{-g}G^{\mu\nu},$$ with $G^{\mu\nu}=R^{\mu\nu}-Rg^{\mu\nu}/2$ being the Einstein tensor, the variation of the action can then be expressed as 

\begin{equation}
\delta S_G=-\frac{1}{8\pi}\int_M\sqrt{-g}\xi_\nu\nabla_\mu G^{\mu\nu}d^4x.
\end{equation}
Since the Einstein-Hilbert action is known to be invariant under diffeomorphisms generated by arbitrary $\xi^\mu$, this is, under an arbitrary element of the diffeomorphism group Diff($M$), then $\nabla_\mu G^{\mu\nu}=0$ has to hold. This is the contracted Bianchi identity, which from this viewpoint, can be seen as a requirement that comes from the covariance of the Einstein-Hilbert action\footnote{A similar argument can be applied to the case of gravity coupled to matter, where we now have that $S[g_{\mu\nu}, \psi_m]=S_G[g_{\mu\nu}]+S_m[g_{\mu\nu}, \psi_m]$. Here $S_m$ is the action functional of a matter field $\psi_m$. Since the theory in the vacuum is covariant, i.e., $\delta S_G=0$, then if the full coupled action is covariant as well, then $\delta S_m=0$. Assuming the matter field equations are satisfied ($\delta S_m/\delta\psi_m=0$), this leads to $$\nabla_\mu T^{\mu\nu}=0, \quad \text{with:} \quad T^{\mu\nu}=\frac{1}{\sqrt{-g}}\frac{\delta S_m}{\delta g_{\mu\nu}},$$ as a consequence of covariance \cite{wald}. Physically, as it is well known, this represents the conservation of the energy-momentum tensor $T^{\mu\nu}$.}. 

On the other hand, in Hamiltonian formulations of GR in which spacetime is foliated by a family of spacelike hypersufaces $\Sigma_t$, and thus can be interpreted as separated into space and time (e.g., those based on ADM analysis), the discussed invariant property is more subtle due to them not being manifestly covariant. In such formulations the total Hamiltonian $H_T$ ends up being a sum of constraints,

\begin{equation}
H_T=H[N]+C[N^a], \quad H[N]=\int_\Sigma N\mathcal{H}d^3x, \quad C[N^a]=\int_\Sigma N^a\mathcal{C}_ad^3x.
\label{HT}
\end{equation}
In equation (\ref{HT}) coordinates on a given spacelike hypersuface $\Sigma$ are denoted by $x^a$ with $a=1,2,3$. Additionally, $H[N]$ and $C[N^a]$ are the so-called Hamiltonian and diffeomorphism constraints, respectively, while $N$ and $N^a$ are the lapse function and the shift vector, who act as Lagrange multipliers. These constraints can be interpreted as generators of gauge transformations. The Hamiltonian constraint is related to the invariance under temporal reparametrizations, and the diffeormorhpism constraint, to invariance under diffeomorphisms on the spacelike hypersuface $\Sigma$. Here, the explicit forms for $\mathcal{H}$ and $\mathcal{C}_a$, which should be understood as densities of their corresponding constraints, will not be given explicitly due to their expressions not being of main interest at the moment (see, for instance, \cite{Bojowald_2010} for a detailed analysis). Instead, we shall focus on the algebra they form under Poisson brackets, this operation is defined on the phase space described by suitable canonical variables.

The mentioned constraint algebra is the following,

\begin{equation}
\{H[N],H[M]\}=-C[q^{ab}(N\partial_bM-M\partial_bN)], \quad \{C[N^a],H[N]\}=H[\pounds_{N^a}N], \quad \{C[N^a],C[M^b]\}=C[\pounds_{N^a}M^b],
\label{CA}
\end{equation}
where $q^{ab}$ is the inverse of the induced spatial metric on $\Sigma$. Note that precisely because of the appearance of this phase space dependent tensor in the result of the $\{H[N],H[M]\}$ bracket, the quantity $q^{ab}(N\partial_bM-M\partial_bN)$ is a structure function, rather than a structure constant, and the constraint algebra does not form a Lie algebra. However, the fact that the Poisson bracket between any two given constraints vanishes on-shell (on the constraint surface) indicates that the constraints are first class and thus, their Hamiltonian flow is tangent to the constraint surface. Put another way, gauge transformations generated by $H[N]$ and $C[N^a]$ do not leave the constraint surface. The algebra shown in (\ref{CA}) is also called the hypersurface deformation algebra. 

So far, the group Diff($M$) and the constraint algebra have been briefly introduced. They are indeed related, but also have fundamental differences between one another. To begin with, the group Diff($M$) does form a Lie algebra consisting of vector fields on $M$ under the Lie bracket operation. In contrast, the constraint algebra (\ref{CA}) fails to be a Lie algebra due to the presence of a structure function. This was to be expected based on the intuition that the foliation chosen to decompose the spacetime would affect its constraint algebra and hence, it naturally depends on the induced metric. This already is a sign that both objects are not identical. In fact, the constraints $H[N]$ and $C[N^a]$ generate general spacetime diffeomorphisms only on-shell, while their algebra can be considered as an off-shell symmetry of the theory in phase space \cite{Thiemann_2007}. Thus, general covariance is encoded into the hypersuface deformation algebra.

\section{III. A Different Approach on Diffeomorphism Invariance in the Hamiltonian Formulation}

While the constraint algebra already contains information regarding the diffeomorphism invariance of a given Hamiltonian written as a sum of contraints, in this paper we wish to provide another tool to identify said property. Specifically, in the past section we showed that if an action is generally covariant, then consistency with the variational techniques demands that

\begin{equation}
\nabla_\mu\mathscr{G}^{\mu\nu}=0, \quad \text{with:} \quad \mathscr{G}^{\mu\nu}=\frac{1}{\sqrt{-g}}\frac{\delta S}{\delta g_{\mu\nu}}.
\label{diffG}
\end{equation}
We will now let $S$ be an arbitrary action. It is clear that for the Einstein-Hilbert action, i.e., when $S=S_G$ as given by equation (\ref{EHS}), we have that $\mathscr{G}^{\mu\nu}=G^{\mu\nu}$ is the Einstein tensor and (\ref{diffG}) immediately holds, avoiding thus a contradiction with invariance under diffeomorphisms. Equation (\ref{diffG}) can be seen as a compatibility criterion. It is guaranteed to hold in the classical theory, but not necessarily in a canonically formulated effective model. The idea is readily applicable to a given action functional $S[g_{\mu\nu}]$ within the Lagrangian framework of any theory of interest. However, in situations when the Hamiltonian in terms of a set of phase space variables is know, instead of a Lagrangian in configuration space, some prior mathematical manipulations are needed before verifying if equation (\ref{diffG}) is satisfied. This is the case, for instance, in effective models found by applying the analysis techniques developed by LQG to symmetry reduced spacetimes. 

Our first step will then be to express $\mathscr{G}^{\mu\nu}$ in terms of the typical quantities that characterize the canonical formulation, these are, phase space variables, Poisson brackets, etc. Since the field equations yielded by both Lagrangian and Hamiltonian formulations are equivalent, this seems as a simple enough task to achieve. Consider a generic phase space described by $\mathbf{\Gamma}=(Q_A;P^A)$. Here, the label $A$ runs through the values $A=1,2,\ldots,n$, where $n$ is the number of degrees of freedom that characterize the spatial part of the gravitational model. Since there are at most six independent coefficients in the induced metric on $\Sigma$, it is obvious that $n\leq6$. This number is complemented by the degrees of freedom provided by the lapse function $N$ and shift vector $N^a$, whose conjugate momenta vanish, to recover a maximum of ten independent metric coefficients. To name a pair of relevant examples, we have that for the metric variables of the ADM analysis, the phase space is given by $$\mathbf{\Gamma}=(q_{ab};\Pi^{ab}), \quad Q_A\rightarrow q_{ab}, \quad P^A\rightarrow \pi^{ab},$$ where the canonical momentum $\pi^{ab}$ is conjugate to the spatial metric $q_{ab}$ and is related to extrinsic curvature. For the Ashtekar-Barbero variables\footnote{It might be worth mentioning that, when expressing the Hamiltonian in Ashtekar-Barbero variables, an additional constraint generally appears. This is, $H_T=H[N]+C[N^a]+G[\lambda_i]$, where $G[\lambda_i]$ is known as the Gauss constraint and $i=1,2,3$ is an internal space index. The Gauss constraint is responsible for generating $SU(2)$ gauge transformations in such a space. Despite this, one can always solve the Gauss constraint explicitly by adequately fixing the gauge and then work in a reduced phase space. In what follows we will assume that this process has been carried out, if needed.} which will be used throughout this paper, we have $$\mathbf{\Gamma}=(A_a^i;E_i^a), \quad Q_A\rightarrow A_a^i, \quad P^A\rightarrow E_i^a.$$ In this case the configuration variables are the components of the so-called Ashtekar connection $A_a^i$ and their conjugate momenta are the densitized triads $E_i^a$ of the spatial metric. With these examples in mind we note that the spacetime metric $g_{\mu\nu}$ only depends on one half of the previous phase spaces, be it configuration variables or their corresponding conjugate momenta. The Lagrange multipliers $N$ and $N^a$ also appear in said metric as a consequence of the foliation employed to split into space and time. Therefore, in what follows we will assume that $g_{\mu\nu}=g_{\mu\nu}(N, N^a, P^A)$. This implies that

\begin{equation}
\frac{\delta S}{\delta N}=\frac{\delta S}{\delta g_{\mu\nu}}\frac{\partial g_{\mu\nu}}{\partial N}, \quad \frac{\delta S}{\delta N^a}=\frac{\delta S}{\delta g_{\mu\nu}}\frac{\partial g_{\mu\nu}}{\partial N^a}, \quad \frac{\delta S}{\delta P^A}=\frac{\delta S}{\delta g_{\mu\nu}}\frac{\partial g_{\mu\nu}}{\partial P^A}.
\label{dSdN}
\end{equation}
Note that we chose the metric to depend on the conjugate momenta $P^A$. As mentioned previously, instead of $P^A$, the configuration variables $Q_A$ could have also been used for this purpose. In the subsequent analysis one can interchange $Q_A\leftrightarrow P^A$ without modifying the end result.

Consider the Hamiltonian density $\mathcal{H}_T=N\mathcal{H}+N^a\mathcal{C}_a$ as the Legendre transformation of the Lagrangian density $\mathcal{L}$, $$\mathcal{H}_T=\dot{Q}_AP^A-\mathcal{L}.$$ A dot denotes the Lie derivative along the vector field $t^\mu$ which indicates the flow of time, i.e., $\dot{Q}_A=\pounds_tQ_A$. One must be aware, though, that a proper computation of $\dot{Q}_A$ requires knowledge of the geometric structure (the rank of the tensor field) that $Q_A$ represents. It is readily seen that, 

\begin{equation}
\frac{\delta S}{\delta N}=-\mathcal{H}, \quad \frac{\delta S}{\delta N^a}=-\mathcal{C}_a, \quad \frac{\delta S}{\delta P^A}=\dot{Q}_A-\{Q_A,H_T\}.
\end{equation}
The desired Hamiltonian related quantities such as constraints and Poisson brackets are already appearing in our expressions. At this point it is convenient to introduce a space of metric variables with the following notation, $$\vec{\Gamma}_P=(N,N^a,P^A),$$ such that the equations in (\ref{dSdN}) can be compactly rewritten as 

\begin{equation}
\frac{\delta S}{\delta\vec{\Gamma}_P}=\frac{\delta S}{\delta g_{\mu\nu}}\frac{\delta g_{\mu\nu}}{\delta\vec{\Gamma}_P}.
\end{equation}
The past expression can be seen as a linear operation that transforms $\vec{S}_g=\delta S/\delta g_{\mu\nu}$ into $\vec{S}_\Gamma=\delta S/\delta\vec{\Gamma}_P$ through the operator $\mathbf{g}_\Gamma=\delta g_{\mu\nu}/\delta\vec{\Gamma}_P$. In more explicit terms,

\begin{equation}
\vec{S}_\Gamma=\mathbf{g}_\Gamma\vec{S}_g,
\label{SGam}
\end{equation}
with
\begin{equation}
    \vec{S}_g=\begin{bmatrix}
           \delta S/\delta g_{00} \\
           \delta S/\delta g_{01} \\
           \vdots \\
					 \delta S/\delta g_{23} \\
           \delta S/\delta g_{33}
         \end{bmatrix}, \quad
		\vec{S}_\Gamma=\begin{bmatrix}
           \delta S/\delta N \\
           \delta S/\delta N^a \\
           \delta S/\delta P^A
         \end{bmatrix}, \quad
		\mathbf{g}_\Gamma=\begin{bmatrix}
		\delta g_{00}/\delta N & 2\delta g_{01}/\delta N & \hdots & \delta g_{23}/\delta N & \delta g_{33}/\delta N \\
	\delta g_{00}/\delta N^a & 2\delta g_{01}/\delta N^a & \hdots & \delta g_{23}/\delta N^a & \delta g_{33}/\delta N^a\\
	\delta g_{00}/\delta P^A & 2\delta g_{01}/\delta P^A & \hdots & \delta g_{23}/\delta P^A & \delta g_{33}/\delta P^A	
\end{bmatrix}.
\nonumber
  \end{equation}
The manner in which these vectors and the matrix $\mathbf{g}_\Gamma$ were constructed already takes into account the symmetry of $g_{\mu\nu}$. This means that, because only the independent components of the metric are being included, both vectors are of dimension $n=10$ and $\mathbf{g}_\Gamma$ is a $10\times10$ matrix. Notice also that due to the mentioned symmetry, a factor of $2$ multiplies the functional derivatives that involve components of $g_{\mu\nu}$ with $\mu\neq\nu$ in $\mathbf{g}_\Gamma$. Without this mathematical manipulation, one ends up with a singular $16\times16$ matrix as a consequence of repeated columns. 

Equation (\ref{SGam}) can be readily inverted to obtain 

\begin{equation}
\vec{\mathscr{G}}=\frac{1}{\sqrt{-g}}\mathbf{g}^{-1}_\Gamma\vec{S}_\Gamma, \quad \vec{\mathscr{G}}=\begin{bmatrix}
           \mathscr{G}^{00} \\
           \mathscr{G}^{01} \\
           \vdots \\
					 \mathscr{G}^{23} \\
           \mathscr{G}^{33}
         \end{bmatrix},
				\label{SG}
\end{equation}
With the entries of the vector $\vec{\mathscr{G}}$ now available, and containing the necessary components of $\mathscr{G}^{\mu\nu}$, the computation of the quantity of interest $\nabla_\mu\mathscr{G}^{\mu\nu}$ can be carried out. 
In what follows we will apply this procedure to effective spherically symmetric models that possess quantum corrections based on LQG. The analogous effective Kantowski-Sachs model, which describes an anisotropic but homogenous cosmology, will also be considered. 

\section{IV. The Spherically Symmetric Model}

Our first objective will be to analyze the general covariance of effective spherically symmetric models with corrections due to Loop Quantum Gravity. Before this, though, let us introduce the classical description in the Hamiltonian formulation for this type of spacetimes. Here only a brief initial summary will be presented, helpful references for an in-depth discussion of the subject are \cite{Bojowald2004, Bojowald2006, Campiglia2007}.

\subsection{IV.A. Classical Description}

Using canonically transformed Ashtekar-Barbero variables, the phase space of the spherically symmetric model is characterized by two degrees of freedom, this is, $\mathbf{\Gamma}=(K_r,K_\varphi ; E^r, E^\varphi)$. In this case $K_r$ and $K_\varphi$ are components of the extrinsic curvature of the spatial hypersurfaces, while $E^r$ and $E^\varphi$, their respective conjugate momenta, are components of the densitized triad of the spatial metric. This information is then complemented by the lapse function $N$, and the only non-trivial component $N_r$ of the shift vector, to fully describe the geometry of the spacetime. In spherical coordinates $\{t,r,\theta,\varphi\}$ the line element can be written as

\begin{equation}
ds^2=-N^2dt^2+\frac{(E^\varphi)^2}{E^r}\left(dr+N_rdt\right)^2+E^rd\Omega^2,
\label{ds2Esf}
\end{equation}
where $d\Omega^2$ is the line element of the two-sphere. We will consider the most general spherically symmetric model, and thus, all of the phase space variables and Lagrange multipliers depend on the chosen time and radial coordinates, e.g., $N=N(t,r)$, $E^\varphi=E^\varphi(t,r)$, $K_\varphi=K_\varphi(t,r)$, etc. The time dependence in the past variables is meant to be understood as their evolution along the spatial hypersurfaces $\Sigma_t$ of the foliation. The Poisson brackets between these canonical variables are given by \cite{Campiglia2007}\footnote{It may be worth mentioning that our notation varies with respect to \cite{Campiglia2007}. The configuration variables $K_r$ and $K_\varphi$ used here correspond respectively to $A_x$ and $\bar{A}_\varphi$ in said reference.}

\begin{equation}
\{K_r(t,r),E^r(t,r')\}=2\gamma\delta(r,r'), \quad \{K_\varphi(t,r),E^\varphi(t,r')\}=2\gamma\delta(r,r').
\end{equation}
Here $\gamma$ is the Barbero-Immirzi parameter. To lighten the notation the coordinate dependence will be omitted hereafter. The Hamiltonian is a sum of two constraints\footnote{As mentioned in footnote 2, during the analysis with Ashtekar-Barbero variables, a non-trivial component of the Gauss constraint would appear as an extra term in the total Hamiltonian. In the expressions (\ref{Dc}) and (\ref{Hc}) for the Hamiltonian and diffeormorphism constraints, this Gauss component has already been solved, thus reducing the phase space to that which was previously introduced, $\mathbf{\Gamma}=(K_r,K_\varphi ; E^r, E^\varphi)$.} $H_T[N,N_r]=H_c[N]+C_c[N_r]$, whose explicit expressions are,

\begin{eqnarray}
C_c[N_r]&=&\int drN_r\left[(E^r)'K_r-E^\varphi K'_\varphi\right], \label{Dc} \\
H_c[N]&=&-\int drN\left[\frac{K_\varphi}{2\gamma^2\sqrt{|E^r|}}\left(4K_rE^r+K_\varphi E^\varphi\right)+\frac{2E^\varphi}{\sqrt{|E^r|}}(1-\Gamma_\varphi^2)+4\sqrt{|E^r|}\Gamma'_\varphi\right],
\label{Hc}
\end{eqnarray}
with $\Gamma_\varphi=-E^{r\prime}/2E^\varphi$. Additionally, a prime denotes derivation with respect to $r$. The algebra of constraints that (\ref{Dc}) and (\ref{Hc}) form under Poisson brackets is

\begin{align}
\{H_c[N],C_c[N_r]\}=&H_c[N_rN'], \quad \{C_c[N_r],C_c[M_r]\}=-C_c[M_rN'_r-N_rM'_r], \nonumber \\
\{H_c[N],H_c[M]\}=&C_c\left[\frac{E^r}{(E^\varphi)^2}\left(MN'-NM'\right)\right].
\label{CAEsf}
\end{align}

To illustrate the use of equation (\ref{SG}), the diffeomorphism invariance of this symmetry reduced model can be verified. We start by specifying our space of metric variables $\vec{\Gamma}_P=(N,N_r,E^r,E^\varphi)$. With this, the vector $\vec{S}_\Gamma$, as well as the matrix $\mathbf{g}_\Gamma$ can be constructed. Note that since there are four independent metric coefficients which, of course, consistently coincide with the two degrees of freedom in phase space added to the lapse function and the radial component of the shift vector, these objects will have dimension $n=4$. Writing this out in explicit terms yields

\begin{eqnarray}
		\vec{S}_\Gamma=\begin{bmatrix}
           \delta S/\delta N \\
           \delta S/\delta N_r \\
           \delta S/\delta E^r \\
					 \delta S/\delta E^\varphi
         \end{bmatrix}&=&
		\begin{bmatrix}
           -\mathcal{H}_c \\
           -\mathcal{C}_c \\
           \dot{K}_r-\{K_r,H_T[N,N_r]\} \\
					 \dot{K}_\varphi-\{K_\varphi,H_T[N,N_r]\}
         \end{bmatrix}, \label{Syg} \\		
		\mathbf{g}_\Gamma=\begin{bmatrix}
		\delta g_{00}/\delta N & 2\delta g_{01}/\delta N & \delta g_{11}/\delta N & 2\delta g_{22}/\delta N \\
	\delta g_{00}/\delta N_r & 2\delta g_{01}/\delta N_r & \delta g_{11}/\delta N_r & 2\delta g_{22}/\delta N_r \\
	\delta g_{00}/\delta E^r & 2\delta g_{01}/\delta E^r & \delta g_{11}/\delta E^r & 2\delta g_{22}/\delta E^r \\
	\delta g_{00}/\delta E^\varphi & 2\delta g_{01}/\delta E^\varphi & \delta g_{11}/\delta E^\varphi & 2\delta g_{22}/\delta E^\varphi	
\end{bmatrix}&=&
		\begin{bmatrix}
		-2N & 0 & 0 & 0 \\
	2N_r(E^\varphi)^2/E^r & 2(E^\varphi)^2/E^r & 0 & 0 \\
	-(N_rE^\varphi/E^r)^2 & -2N_r(E^\varphi/E^r)^2 & -(E^\varphi/E^r)^2 & 2	\\
	2N_r^2E^\varphi/E^r & 4N_rE^\varphi/E^r & 2E^\varphi/E^r & 0
\end{bmatrix},
\nonumber
\end{eqnarray}
where the last column of $\mathbf{g}_\Gamma$ appears multiplied by $2$ due to the fact that $g_{33}=g_{22}\sin^2\theta$. After inverting $\mathbf{g}_\Gamma$ we can find the components of what in this case can be considered as the Einstein tensor in phase space, 

\begin{align}
    \mathscr{G}^{00}&=\frac{-\mathcal{H}_c}{2N^2E^\varphi\sqrt{E^r}}, \quad \mathscr{G}^{01}=\frac{-1}{NE^\varphi\sqrt{E^r}}\left(\frac{E^r\mathcal{C}_c}{\gamma(E^\varphi)^2}-\frac{N_r\mathcal{H}_c}{2N}\right), \nonumber \\
		\mathscr{G}^{11}&=\frac{\sqrt{E^r}}{2N(E^\varphi)^2}\left(\{K_\varphi,H_T\}-\dot{K}_\varphi\right)-\frac{N_r}{NE^\varphi\sqrt{E^r}}\left(\frac{N_r\mathcal{H}_c}{2N}-\frac{2E^r\mathcal{C}_c}{\gamma(E^\varphi)^2}\right), \label{GEsf} \\ 
		\mathscr{G}^{22}&=\frac{1}{2NE^\varphi\sqrt{E^r}}\left(\{K_r,H_T\}-\dot{K}_r\right)+\frac{1}{4N(E^r)^{3/2}}\left(\{K_\varphi,H_T\}-\dot{K}_\varphi\right). \nonumber
\end{align}
The missing non-vanishing component of $\mathscr{G}^{\mu\nu}$ is given by the spherical symmetry of this particular spacetime, this is, $\mathscr{G}^{33}=\mathscr{G}^{22}/\sin^2\theta$. When the set of expressions found in (\ref{GEsf}) is evaluated in solutions of the Hamilton equations, $\dot{E}^r=\{E^r,H_T[N,N_r]\}$ and $\dot{E}^\varphi=\{E^\varphi,H_T[N,N_r]\}$ which define the extrinsic curvature variables, i.e.,

\begin{equation}
    K_\varphi=\frac{\gamma}{N\sqrt{E^r}}\left[\dot{E}^r-N_r(E^r)'\right], \quad K_r=\frac{\gamma}{2N(E^r)^{3/2}}\left(E^\varphi[N_r(E^r)'-\dot{E}^r]+2E^r[\dot{E}^\varphi-(N_rE^\varphi)']\right),
\label{K}
\end{equation}
they reduce exactly to $G^{\mu\nu}$ as computed by using only the metric in (\ref{ds2Esf}). We will denote the process of evaluating in solutions of this half of the Hamilton equations as $\mathscr{G}^{\mu\nu}|_K$. Hence, in the classical case we have that $\mathscr{G}^{\mu\nu}|_K=G^{\mu\nu}$. This justifies calling the components in (\ref{GEsf}) as the Einstein tensor in phase space. As it is well-known, the other half of Hamilton equations, $\dot{K}_r=\{K_r,H_T[N,N_r]\}$ and $\dot{K}_\varphi=\{K_\varphi,H_T[N,N_r]\}$, along with (\ref{K}) and the constraints $H_c=C_c=0$ are equivalent to the Einstein field equations in vacuum $G^{\mu\nu}=0$. However, as it can be easily seen from this example, the constraints and the Poisson brackets, $\{K_r,H_T\}$ and $\{K_\varphi,H_T\}$ alone, do not coincide with the Einstein tensor. Some further manipulation is needed. This procedure provides the analogous expressions for $G^{\mu\nu}$ in the classical phase space.

Using now the only non-zero components of (\ref{GEsf}), it is straightforward (though somewhat tedious) to calculate 

\begin{equation}
\nabla_\mu\mathscr{G}^{\mu\nu}=\sum_{i=0}^2\left[A_{(i)}^\nu\partial^i_r\left(\dot{E}^r-\{E^r,H_T[N,N_r]\}\right)+B_{(i)}^\nu\partial^i_r\left(\dot{E}^\varphi-\{E^\varphi,H_T[N,N_r]\}\right)\right].
\label{DGuv}
\end{equation}
Here, $A_{(i)}^\mu=A_{(i)}^\mu(\mathbf{\Gamma}_+)$ and $B_{(i)}^\mu=A_{(i)}^\mu(\mathbf{\Gamma}_+)$ are spacetime vectors that depend on the variables of phase space $\mathbf{\Gamma}$ and on the Lagrange multipliers $N$ and $N_r$, i.e., $\mathbf{\Gamma}_+=(N,N_r,\mathbf{\Gamma})$. For compactness, the explicit expressions of $A_{(i)}^\mu$ and $B_{(i)}^\mu$ are not shown at this point due to them being quite long. Nonetheless, for the sake of completeness, appendix A will include these complementary calculations as reference. Also, the $(i)$ indexes should be understood as labels, they do not represent components in any kind of space. It is clear that when the Hamilton equations that lead to (\ref{K}) hold, then $$\nabla_\mu\mathscr{G}^{\mu\nu}|_K=0.$$ Therefore, the classical spherically symmetric reduced action implied by (\ref{Dc}) and (\ref{Hc}) is generally covariant. To be specific, we will say that the action is covariant with respect to a given metric, in this case, the classical line element given by (\ref{ds2Esf}). This result was, of course, expected. Note that the constraints or the other half of the Hamilton equations, i.e., $\dot{E}^r=\{E^r,H_T[N,N_r]\}$ and $\dot{E}^\varphi=\{E^\varphi,H_T[N,N_r]\}$, were not required to be satisfied during the previous analysis.

Having discussed the classical description of the spherically symmetric model, we now consider its quantum counterpart. The classical algebra (\ref{CAEsf}) is in general altered when one incorporates quantum corrections to the total Hamiltonian, seemingly implying the breakdown of general covariance. As mentioned earlier, we will focus on two types of corrections that normally appear in effective models yielded by LQG, namely, inverse triad and holonomy corrections. As a first approach, and following \cite{Tibrewala_2012, Tibrewala_2014}, these corrections will be treated separately.

\subsection{IV.B. Inverse Triad Corrections} 

In the process of finding a suitable Hamiltonian constraint operator $\hat{H}_c$ during loop quantization, issues arise when dealing with a factor of $q^{-1/2}$ ($q=\det[q_{ab}]$) that generally appears in the classical expression of $H_c$. The problem lies in the fact that the straightforward promotion to an operator $\hat{q}^{-1/2}$ of the mentioned classical quantity does not correspond to a densely defined operator in the kinematical Hilbert space $\mathcal{H}_{kin}$. To solve this problem, regularization procedures have been proposed, the most well-known due to Thiemann. It consists in replacing the troublesome factor with an appropriate Poisson bracket between the classical volume and holonomies along certain directions (see section 6.3 of \cite{AshtekarLewandowski} and references therein). Following the Dirac quantization procedure, those Poisson brackets are in turn promoted to commutators, whose presence add non-trivial information about the quantum dynamics of the geometry in the effective model. These are the so-called inverse triad corrections. 

For the case of spherical symmetry, the simplest inverse triad corrections are obtained by replacing $$\frac{1}{\sqrt{E^r}}\rightarrow\frac{\alpha_1(E^r)}{\sqrt{E^r}} \quad \text{and} \quad \sqrt{E^r}\rightarrow\alpha_2(E^r)\sqrt{E^r}$$ in the classical Hamiltonian constraint. The explicit form of the correction functions $\alpha_{1,2}(E^r)$ are determined depending on the schemed followed to find the effective model. Nevertheless, for the remainder of this paper we will treat them as unspecified functions. This yields an effective version of equation (\ref{Hc}) while leaving the diffeomorphism constraint unchanged, namely,

\begin{eqnarray}
C_{eff}[N_r]&=&C_c[N_r], \nonumber \\
H_{eff}[N]&=&-\int drN\left[\frac{K_\varphi}{2\gamma^2\sqrt{|E^r|}}\left(4\alpha_2(E^r)K_rE^r+\alpha_1(E^r)K_\varphi E^\varphi\right)+\frac{2\alpha_1(E^r)E^\varphi}{\sqrt{|E^r|}}(1-\Gamma_\varphi^2)+4\alpha_2(E^r)\sqrt{|E^r|}\Gamma'_\varphi\right].
\label{HeffInv}
\end{eqnarray}
The total Hamiltonian is naturally $H_T^{eff}[N,N_r]=H_{eff}[N]+C_{eff}[N_r]$. Finally, the effective model reduces to the classical case when $\alpha_1=\alpha_2=1$.

With these effective constraints defined we can now proceed to test covariance by the method described in the past section. However, a specific metric must be chosen first for this purpose. Making the apparently obvious election, that is, using the line element (\ref{ds2Esf}), leads to the loss of diffeomorphism invariance. In fact, equations (\ref{GEsf}) are still valid for this analysis as long as the classical constraints are replaced by the effective ones ($\mathcal{H}_c\rightarrow\mathcal{H}_{eff}$ and $\mathcal{C}_c\rightarrow\mathcal{C}_{eff}$). Due to this change, one must keep in mind that those quantities will no longer reduce to the components of the Einstein tensor, but rather to an effective $\mathscr{G}^{\mu\nu}$. Thus $\nabla_\mu\mathscr{G}^{\mu\nu}|_K=0$ is not guaranteed to hold. The Hamilton equations, $$\dot{E}^r=\{E^r,H_T^{eff}[N,N_r]\}, \quad \dot{E}^\varphi=\{E^\varphi,H_T^{eff}[N,N_r]\},$$ now imply

\begin{equation}
    K_\varphi=\frac{\gamma}{\alpha_2N\sqrt{E^r}}\left[\dot{E}^r-N_r(E^r)'\right], \quad K_r=\frac{\gamma}{2\alpha_2N(E^r)^{3/2}}\left(\frac{\alpha_1E^\varphi}{\alpha_2}[N_r(E^r)'-\dot{E}^r]+2E^r[\dot{E}^\varphi-(N_rE^\varphi)']\right),
\label{K2}
\end{equation}
For this effective model, the divergence of $\mathscr{G}^{\mu\nu}$ can now be expressed as

\begin{align}
\nabla_\mu\mathscr{G}^{\mu\nu}=&\sum_{i=0}^2\left[\bar{A}_{(i)}^\nu\partial^i_r\left(\dot{E}^r-\{E^r,H_T^{eff}[N,N_r]\}\right)+\bar{B}_{(i)}^\nu\partial^i_r\left(\dot{E}^\varphi-\{E^\varphi,H_T^{eff}[N,N_r]\}\right)\right] \nonumber \\
&+(\alpha_2^2-1)f_{(1)}^\nu+\frac{d\alpha_2}{dE^r}f_{(2)}^\nu.
\label{DG1}
\end{align}
In contrast with (\ref{DGuv}), $\nabla_\mu\mathscr{G}^{\mu\nu}$ acquires two additional terms, where $f_{(1,2)}^\mu=f_{(1,2)}^\mu(\mathbf{\Gamma}_+)$ are again complicated vectors that are not shown explicitly, but can be found in appendix A along with $\bar{A}_{(i)}^\mu$ and $\bar{B}_{(i)}^\mu$. These vectors do not vanish in the constraint surface nor when the Hamilton equations hold. Thus, $\nabla_\mu\mathscr{G}^{\mu\nu}|_K\neq0$. The effective action is then generally covariant with respect to the classical metric only if $\alpha_2^2(E^r)=1$. It is worth highlighting that this will happen independently of the $\alpha_1$ correction function, the only relevant quantity for general covariance is $\alpha_2$. This is no surprise since previous works \cite{Tibrewala_2012, Tibrewala_2014} in this direction have already explored the algebra of constraints for such effective models, finding thereby

 \begin{equation}
        \{H_{eff}[N],H_{eff}[M]\}=C_{eff}\left[\frac{\alpha_2^2E^r}{(E^\varphi)^2}\left(MN'-NM'\right)\right].
				\label{HHeff}
    \end{equation}
The rest of the Poisson brackets involving the effective constraints are unmodified. Despite this, because of the presence of an additional $\alpha_2^2$ factor, this algebra is said to be deformed (compare to (\ref{CAEsf})). Likewise, $\alpha_1$ does not appear in equation (\ref{HHeff}) and when $\alpha_2^2=1$, the usual hypersurface deformation algebra of the spherically symmetric model is recovered. The result in (\ref{DG1}) is therefore consistent with the previously reported effective algebra.

In light of this deformed algebra of effective constraints, Tibrewala has already proposed alternative ways to recover diffeomorphism invariance in these modified models. The basic idea is that, in the case of a deformed algebra, general covariance can be restored by mapping the phase space variables to some auxiliary variables that modify the spacetime metric $g_{\mu\nu}\rightarrow\bar{g}_{\mu\nu}$ such that general covariance is restored. This new metric is sometimes referred to as an ``emergent'' metric. Throughout the paper all of this type of metrics will be denoted by a bar. As a manner of example, the Poisson bracket (\ref{HHeff}) strongly suggests to make the change $$E^r\rightarrow\bar{E}^r=\alpha_2^2E^r, \quad E^\varphi\rightarrow E^\varphi.$$ This yields a new metric 

\begin{equation}
\bar{ds}^2_{(1)}=\bar{g}_{\mu\nu}^{(1)}dx^\mu dx^\nu=-N^2dt^2+\frac{(E^\varphi)^2}{\bar{E}^r}\left(dr+N_rdt\right)^2+\bar{E}^rd\Omega^2,
\label{ds2Esf2}
\end{equation}
so that the classical constraint algebra structure is recovered $$\{H_{eff}[N],H_{eff}[M]\}=C_{eff}\left[\frac{\bar{E}^r}{(E^\varphi)^2}\left(MN'-NM'\right)\right].$$
It is important to make it clear that the map $E^r\rightarrow\bar{E}^r$ is meant to change only the line element of the model. The effective Hamiltonian $H_T^{eff}$ is left unaltered. 

In \cite{Tibrewala_2014} the effective equations of motion given by $H_T^{eff}$ were solved. Inserting those particular solutions in a line element of the previous form, it turned out that said metric was diffeomorphism invariant. This was proven by verifying that solutions of the equations of motion were mapped to solutions when performing a coordinate transformation in $\bar{g}_{\mu\nu}^{(1)}$. In what follows we will prove that indeed covariance can be generally restored for this model by utilizing the test presented in this paper.

The procedure is completely analogous to that already described in the past subsection for the classical spherically symmetric model. The use of the variables $\vec{\Gamma}_P=(N,N_r,E^r,E^\varphi)$ is maintained. However, the modified metric $\bar{g}_{\mu\nu}^{(1)}$ will now be utilized instead of the initial $g_{\mu\nu}$. For the vector $\vec{S}_\Gamma$ and the matrix $\mathbf{g}_\Gamma$ we obtain

\begin{eqnarray}
		\vec{S}_\Gamma=&=&
		\begin{bmatrix}
           -\mathcal{H}_{eff} \\
           -\mathcal{C}_{eff} \\
           \dot{K}_r-\{K_r,H_T^{eff}[N,N_r]\} \\
					 \dot{K}_\varphi-\{K_\varphi,H_T^{eff}[N,N_r]\}
         \end{bmatrix}, \nonumber \\		
		\mathbf{g}_\Gamma=&=&
		\begin{bmatrix}
		-2N & 0 & 0 & 0 \\
	2N_r(E^\varphi)^2/\bar{E}^r & 2(E^\varphi)^2/\bar{E}^r & 0 & 0 \\
	(N_rE^\varphi)^2d\left[(\bar{E}^r)^{-1}\right]/dE^r & 2N_r(E^\varphi)^2d\left[(\bar{E}^r)^{-1}\right]/dE^r & (E^\varphi)^2d\left[(\bar{E}^r)^{-1}\right]/dE^r & 2d\bar{E}^r/dE^r	\\
	2N_r^2E^\varphi/\bar{E}^r & 4N_rE^\varphi/\bar{E}^r & 2E^\varphi/\bar{E}^r & 0
\end{bmatrix}.
\nonumber
\end{eqnarray}

As in the past example, the components of the effective Einstein tensor can be found via the inversion of this $\mathbf{g}_\Gamma$ matrix. We thus have

\begin{align}
    \mathscr{G}^{00}&=\frac{-\mathcal{H}_{eff}}{2N^2E^\varphi\sqrt{\bar{E}^r}}, \quad \mathscr{G}^{01}=\frac{-1}{NE^\varphi\sqrt{\bar{E}^r}}\left(\frac{\bar{E}^r\mathcal{C}_{eff}}{\gamma(E^\varphi)^2}-\frac{N_r\mathcal{H}_{eff}}{2N}\right), \nonumber \\
		\mathscr{G}^{11}&=\frac{\sqrt{\bar{E}^r}}{2N(E^\varphi)^2}\left(\{K_\varphi,H_T^{eff}\}-\dot{K}_\varphi\right)-\frac{N_r}{NE^\varphi\sqrt{\bar{E}^r}}\left(\frac{N_r\mathcal{H}_{eff}}{2N}-\frac{2\bar{E}^r\mathcal{C}_{eff}}{\gamma(E^\varphi)^2}\right), \label{GEsf2} \\ 
		\mathscr{G}^{22}&=\frac{1}{2NE^\varphi\sqrt{\bar{E}^r}d\bar{E}^r/dE^r}\left(\{K_r,H_T^{eff}\}-\dot{K}_r\right)-\frac{\sqrt{\bar{E}^r}}{4Nd\bar{E}^r/dE^r}\frac{d}{dE^r}\left(\frac{1}{\bar{E}^r}\right)\left(\{K_\varphi,H_T^{eff}\}-\dot{K}_\varphi\right), \nonumber
\end{align}
additionally $\mathscr{G}^{33}=\mathscr{G}^{22}/\sin^2\theta$. One can easily verify that if $\alpha_1=\alpha_2=1$, which makes $\bar{E}^r=E^r$, then this effective Einstein tensor reduces to (\ref{GEsf}). It turns out that even for non-trivial $\alpha_{1,2}$, the divergence $\nabla_\mu\mathscr{G}^{\mu\nu}$ can be written as a sum involving the Hamilton equations for $\dot{E}^r$ and $\dot{E}^\varphi$, this is, 

\begin{equation}
\nabla_\mu\mathscr{G}^{\mu\nu}=\sum_{i=0}^2\left[\mathcal{A}_{(i)}^\nu\partial^i_r\left(\dot{E}^r-\{E^r,H_T^{eff}[N,N_r]\}\right)+\mathcal{B}_{(i)}^\nu\partial^i_r\left(\dot{E}^\varphi-\{E^\varphi,H_T^{eff}[N,N_r]\}\right)\right],
\label{DG2}
\end{equation}
where again appendix A contains the complete form of this expression. The last two problematic terms of equation (\ref{DG1}) no longer appear as a result of using a different metric. Therefore we have that $\nabla_\mu\mathscr{G}^{\mu\nu}|_K=0$ and general covariance is regained. Note that for this to be the case, the metric required modifications from the correction functions also present in $H_T^{eff}$. To be precise, the effective model, whose dynamics is described by the constraints in (\ref{HeffInv}), is invariant under diffeomorphisms when the line element of the theory is given by $\bar{g}_{\mu\nu}^{(1)}$ instead of $g_{\mu\nu}$. 

At this point, one may wonder if the metric that restores covariance for these inverse triad corrections is unique. Another proposal to accomplish this was already reported in \cite{Tibrewala_2012}, and it consists in mapping $$N\rightarrow\bar{N}=\alpha_2N,$$ while leaving the rest of the phase space variables and shift vector unchanged so that, 

\begin{equation}
\bar{ds}^2_{(2)}=\bar{g}_{\mu\nu}^{(2)}dx^\mu dx^\nu=-\bar{N}^2dt^2+\frac{(E^\varphi)^2}{E^r}\left(dr+N_rdt\right)^2+E^rd\Omega^2.
\label{ds2Esf3}
\end{equation}
The test for covariance proceeds along the same steps as before, i.e., defining the space of metric variables as $\vec{\Gamma}_P=(N,N_r,E^r,E^\varphi)$ and calculating the components of the effective Einstein tensor through the use of equation (\ref{SG}). For completeness, we show some of the calculations involved in this process,

\begin{align}	
		\mathbf{g}_\Gamma&=
		\begin{bmatrix}
		-2\alpha_2\bar{N} & 0 & 0 & 0 \\
	2N_r(E^\varphi)^2/E^r & 2(E^\varphi)^2/E^r & 0 & 0 \\
	-(N_rE^\varphi/E^r)^2-d(\bar{N}^2)/dE^r & -2N_r(E^\varphi/E^r)^2 & -(E^\varphi/E^r)^2 & 2	\\
	2N_r^2E^\varphi/E^r & 4N_rE^\varphi/E^r & 2E^\varphi/E^r & 0
\end{bmatrix},
\nonumber \\
    \mathscr{G}^{00}&=\frac{-\mathcal{H}_{eff}}{2\alpha_2\bar{N}^2E^\varphi\sqrt{E^r}}, \quad \mathscr{G}^{01}=\frac{-1}{\bar{N}E^\varphi\sqrt{E^r}}\left(\frac{E^r\mathcal{C}_{eff}}{\gamma(E^\varphi)^2}-\frac{N_r\mathcal{H}_{eff}}{2\alpha_2\bar{N}}\right), \nonumber \\
		\mathscr{G}^{11}&=\frac{\sqrt{E^r}}{2\bar{N}(E^\varphi)^2}\left(\{K_\varphi,H_T^{eff}\}-\dot{K}_\varphi\right)-\frac{N_r}{\bar{N}E^\varphi\sqrt{E^r}}\left(\frac{N_r\mathcal{H}_{eff}}{2\alpha_2\bar{N}}-\frac{2E^r\mathcal{C}_{eff}}{\gamma(E^\varphi)^2}\right), \nonumber \\ 
		\mathscr{G}^{22}&=\frac{1}{2\bar{N}E^\varphi\sqrt{E^r}}\left(\{K_r,H_T^{eff}\}-\dot{K}_r\right)+\frac{1}{4\bar{N}(E^r)^{3/2}}\left(\{K_\varphi,H_T^{eff}\}-\dot{K}_\varphi\right)+\frac{\mathcal{H}_{eff}}{2E^\varphi\sqrt{E^r}}\frac{d}{dE^r}\left(\frac{1}{\alpha_2}\right). \nonumber
\end{align}
Similarly, we have that 

\begin{equation}
\nabla_\mu\mathscr{G}^{\mu\nu}=\frac{1}{\alpha_2^2}\sum_{i=0}^2\left[\mathcal{A}_{(i)}^\nu\partial^i_r\left(\dot{E}^r-\{E^r,H_T^{eff}[N,N_r]\}\right)+\mathcal{B}_{(i)}^\nu\partial^i_r\left(\dot{E}^\varphi-\{E^\varphi,H_T^{eff}[N,N_r]\}\right)\right],
\label{DG5}
\end{equation}
which reduces to $\nabla_\mu\mathscr{G}^{\mu\nu}|_K=0$. Diffeomorphism invariance can hence be preserved for the effective model with inverse triad corrections with a different metric other than $\bar{g}_{\mu\nu}^{(1)}$, but still with a correction function present in its coefficients.

The consistency of the above result with the modified algebra of constraints can be easily seen as follows. Define an auxiliary Hamiltonian density $\bar{\mathcal{H}}_{eff}=\mathcal{H}_{eff}/\alpha_2$ such that $H_{eff}[N]=\bar{H}_{eff}[\bar{N}]$. The auxiliary constraint algebra is then

\begin{align}
\{\bar{H}_{eff}[\bar{N}],C_{eff}[N_r]\}=\bar{H}_{eff}[N_r\bar{N}'], \quad \{\bar{H}_{eff}[\bar{N}],\bar{H}_{eff}[\bar{M}]\}=C_{eff}\left[\frac{E^r}{(E^\varphi)^2}\left(\bar{M}\bar{N}'-\bar{N}\bar{M}'\right)\right],
\label{AuxCA}
\end{align}
where $\bar{N}$ is considered as a phase space independent function and naturally the $\{C_{eff}[N_r],C_{eff}[M_r]\}$ bracket does not suffer any changes with respect to (\ref{CAEsf}). The structure of the classical algebra of constraints is thus obtained. Indeed, the auxiliary Hamiltonian constraint $\bar{H}_{eff}$ is equivalent to setting $\alpha_1\rightarrow\alpha_1/\tilde{\alpha}_2$ and $\alpha_2=1$ in $H_{eff}$ of (\ref{HeffInv}). We have already proved that if $\alpha_2=1$, then the model described by $H_T^{eff}$ is generally covariant with respect to the classical metric (\ref{ds2Esf}), see equation (\ref{DG1}).

Finally, due to the spherical symmetry of the spacetime, it can be realized that the only inverse spatial metric coefficient involved in the bracket $\{H[N],H[M]\}$ of equation (\ref{CA}) is $q^{11}=E^r/(E^\varphi)^2$. The lapse function $N$ and the radial component of the shift vector $N_r$ are also part of the constraint algebra as test functions. Therefore, based on the previous calculations, one can restore covariance by altering any of these variables in the line element so long as the classical structure of the algebra is kept. In fact, for this case, $q^{22}$ and $q^{33}=q^{22}/\sin^2\theta$ do not affect covariance. This can be seen by considering the two following metrics

\begin{eqnarray}
\bar{ds}^2&=&\bar{g}_{\mu\nu}^{(IT1)}dx^\mu dx^\nu=-N^2dt^2+\frac{(E^\varphi)^2}{\bar{E}^r}\left(dr+N_rdt\right)^2+q_2(E^r,E^\varphi)d\Omega^2, \label{ds2Inv1} \\
\bar{ds}^2&=&\bar{g}_{\mu\nu}^{(IT2)}dx^\mu dx^\nu=-\bar{N}^2dt^2+\frac{(E^\varphi)^2}{E^r}\left(dr+N_rdt\right)^2+\bar{q}_2(E^r,E^\varphi)d\Omega^2.
\label{ds2Inv2}
\end{eqnarray}
Here, $q_2(E^r,E^\varphi)$ and $\bar{q}_2(E^r,E^\varphi)$ are arbitrary functions of the conjugate momenta. If the covariance test described so far is applied separately to both metrics one can find that $\nabla_\mu\mathscr{G}^{\mu\nu}|_K=0$, i.e., the two of them are diffeomorphism invariant. The steps of the analysis will be omitted to avoid unnecessary repetition. 

Some relevant comments can be made about the previous pair of line elements. Firstly, It is readily seen that when $\bar{q}_2=\alpha_2^2q_2$, these two metrics are related by a conformal transformation, in particular, $$\bar{g}_{\mu\nu}^{(IT2)}=\alpha_2^2\bar{g}_{\mu\nu}^{(IT1)}.$$ They share thus the same causal structure, which is relevant in the case of event horizons for black hole solutions. A natural choice for the $q_2$ function, motivated by the classical spherical model, is simply $q_2=E^r$. One might next wonder if (\ref{ds2Inv1}) and (\ref{ds2Inv2}) are connected through a diffeomorphism, i.e., if $\bar{g}_{\mu\nu}^{(IT2)}=\pounds_\xi\bar{g}_{\mu\nu}^{(IT1)}$. However, it is not possible to find a consistent vector $\xi^\mu$ that generates such a transformation\footnote{Under diffeomorphisms induced by $\xi^\mu=(\xi^0,\xi^1,0,0)$, the relevant metric quantities transform as $$N^2\rightarrow N^2+2N\delta N, \quad N_r\rightarrow N_r+\delta N_r, \quad \bar{E}_1\rightarrow\bar{E}_1+\delta\bar{E}_1, \quad q_2\rightarrow q_2+\delta q_2,$$ with $\bar{E}_1=(E^\varphi)^2/\bar{E}^r$ and 
\begin{align}
\delta N=&(N\xi^0)\dot{}+N'\xi^1-N_rN(\xi^0)', \quad \delta N_r=\dot{\xi}^1+(N_r\xi^0)\dot{}+N_r'\xi_1-N_r\left[(\xi^1)'+N_r(\xi^0)'\right]-\frac{1}{\bar{E}_1}N^2(\xi^0)', \nonumber \\
\delta\bar{E}_1=&\dot{\bar{E}}_1\xi^0+\bar{E}'_1\xi^1+2\bar{E}_1\left[(\xi^1)'+N_r(\xi^0)'\right], \quad \delta q_2=\dot{q}_2\xi^0+q'_2\xi^1. \nonumber
\end{align}
There is no vector $\xi^\mu$ such that $$N^2\rightarrow \bar{N}^2, \quad N_r\rightarrow N_r, \quad \bar{E}_1\rightarrow\frac{(E^\varphi)^2}{E^r}, \quad q_2\rightarrow\bar{q}_2,$$ since it would define an inconsistent system of equations. The above transformations cannot be therefore performed simultaneously.}, consequently the space-times described by both geometries are different. The question regarding which of them should be used as the correct effective metric must be necessarily based on physical arguments as boundary conditions, asymptotic limits, etc., and once explicit solutions of the modified dynamics are known (recall that the field equations do not change depending on the choice of metric). Hence we leave this issue for future work in which specific cases of $\alpha_2$ corrections could be treated. Another point to notice is that if, in an attempt to find other possible covariant line elements, the shift is modified by $N_r\rightarrow\bar{N}_r$, it would lead to the breakdown of the classical algebra. Also, an alteration of the type $E^\varphi\rightarrow E^\varphi/\alpha_2$ can be easily seen to be equivalent to the case already considered in metric (\ref{ds2Inv1}). This exhausts all possible corrections in the metric coefficients and hence, the effective Hamiltonian (\ref{HeffInv}) with inverse triad corrections is generally covariant with respect to only two family of metrics: $\bar{g}_{\mu\nu}^{(IT1)}$ and $\bar{g}_{\mu\nu}^{(IT2)}$, as given by equations (\ref{ds2Inv1}) and (\ref{ds2Inv2}), respectively. 

\subsection{IV.C. Holonomy Corrections}

The other type of corrections that characterize effective Loop Quantum Gravity models are those generated by holonomies. Taking as an example the configuration variables of the spherically symmetric spacetime, these modifications appear because the operators related to extrinsic curvature $\hat{K}_r$ and $\hat{K}_\varphi$ fail to exist in the kinematical Hilbert space of the model. This is a consequence of the discrete property of geometry, which leads to the infinitesimal generator of translations not being well defined. Instead, holonomy operators $\hat{h}_r[K_r]$ and $\hat{h}_{\theta,\varphi}[K_\varphi]$ are employed for the loop quantization procedure. They represent parallel transport along the edges of the preferred coordinate directions $r$, $\theta$ and $\varphi$. 

Here, we will consider only the so-called point-holonomies $$\hat{h}_{\theta,\varphi}[K_\varphi]=\hat{e}^{\frac{i}{4}\delta K_\varphi},$$ where $\delta=\delta(E^r)$ represents the length of the holonomy, which we allow to be phase space dependent. Variable $\delta$ quantization schemes for the spherically symmetric model and analogous to those of LQC \cite{Improved} (sometimes called the $\bar{\mu}$ schemes), have already been proposed in \cite{chiou, GambiniImproved}. They likewise enable control over the holonomy parameter such that it can be made arbitrarily small in the classical regime, avoiding thusly unphysical predictions. Naturally, these quantum properties have also been studied from the effective theory perspective \cite{asier, zhang, belfaqih, Tibrewala_2012, Kelly}. At the effective level, one expects that holonomy features will alter the dynamics of the geometry in a non-trivial way as to incorporate the mentioned discrete aspect of space. An effective Hamiltonian constraint that contains such effects can be written as

\begin{equation}
H_{eff}[N]=-\int drN\left[\frac{1}{2\gamma^2\sqrt{|E^r|}}\left(4K_rE^r\gamma_2(K_\varphi,E^r)+\gamma_1^2(K_\varphi,E^r)E^\varphi\right)+\frac{2E^\varphi}{\sqrt{|E^r|}}(1-\Gamma_\varphi^2)+4\sqrt{|E^r|}\Gamma'_\varphi\right].
\label{HeffHol}
\end{equation}
Again, the diffeomorphism constraint (\ref{Dc}) suffers no modifications. The functions $\gamma_{1,2}(K_\varphi,E^r)$ represent corrections due to holonomies and when $\gamma_1(K_\varphi,E^r)=\gamma_2(K_\varphi,E^r)=K_\varphi$, the constraint reduces to the classical one. The explicit expressions for these functions depend on the scheme used to obtain the effective model, nevertheless, the algebra of effective constraints already imposes some restrictions. As already noted by Tibrewala in \cite{Tibrewala_2012}, an anomalous term appears in the bracket 

\begin{align}
        \{H_{eff}[N],H_{eff}[M]\}=&C_{eff}\left[\frac{E^r}{(E^\varphi)^2}\frac{\partial\gamma_2}{\partial K_\varphi}\left(MN'-NM'\right)\right] \label{HHhol} \\
				&+\int\frac{2(E^r)'}{\gamma^2E^\varphi}\left(\gamma_2-\gamma_1\frac{\partial\gamma_1}{\partial K_\varphi}+2E^r\frac{\partial\gamma_2}{\partial E^r}\right)\left(NM'-MN'\right)dr \nonumber.
    \end{align}
In order for the algebra to be closed, and for the constraints to consequently remain first class, the second line in (\ref{HHhol}) has to vanish. 

Let us begin by analyzing first the case of a constant $\delta$ parameter, hence $\gamma_{1,2}(K_\varphi)$. If one considers the heuristic replacement of $K_{\varphi}\rightarrow\gamma_2(K_\varphi)=\sin(\delta K_\varphi)/\delta$, typical in effective LQG analysis, then the reduced anomaly-free condition, $$\gamma_2-\gamma_1\frac{d\gamma_1}{dK_\varphi}=0,$$ yields $\gamma_1(K_\varphi)=\sqrt{2\left[c-\cos(\delta K_\varphi)\right]}/\delta$. With a convenient choice of integration constant $c=1$, this becomes $\gamma_1=2\sin(\delta K_\varphi/2)/\delta$. These are the explicit forms of the correction functions that will be used for the rest of this section.

Examining the deformed Poisson bracket (\ref{HHhol}), and once the anomalous term has been taken care of, the map $$E^\varphi\rightarrow\bar{E}^\varphi=\frac{E^\varphi}{\sqrt{\left|d\gamma_2/dK_\varphi\right|}},$$ can be proposed as a way to restore general covariance. With the purpose of simplifying calculations, we follow Tibrewala and perform the canonical transformation 

\begin{equation}
\bar{E}^\varphi=\frac{E^\varphi}{\sqrt{\left|d\gamma_2/dK_\varphi\right|}}, \quad \bar{K}_\varphi=\int\sqrt{\left|\frac{d\gamma_2}{dK_\varphi}\right|}dK_\varphi,
\label{phiT}
\end{equation}
where in this case $d\gamma_2/dK_\varphi=\cos(\delta K_\varphi)$. The phase space hence becomes $\mathbf{\Gamma}=(K_r,\bar{K}_\varphi ; E^r, \bar{E}^\varphi)$. Now we need to express the constraints of the Hamiltonian in terms of these new canonical variables, substituting at the same time the specific form proposed for the correction functions $\gamma_{1,2}(K_\varphi)$. This yields, 

\begin{align}
H_{eff}[N]=-\int drN\left[\frac{}{}\right.&\frac{2}{\gamma^2\delta\sqrt{|E^r|}}\left(K_rE^r\sin(\delta K_\varphi)+\frac{1}{\delta^2}\sin^2\left(\frac{\delta}{2}K_\varphi\right)\sqrt{\left|\cos(\delta K_\varphi)\right|}\bar{E}^\varphi\right) \nonumber \\
&\left.+2\bar{E}^\varphi\sqrt{\frac{\left|\cos(\delta K_\varphi)\right|}{|E^r|}}(1-\Gamma_\varphi^2)+4\sqrt{|E^r|}\Gamma'_\varphi\right], \label{HDHol} \\
C_{eff}[N_r]=\int drN_r\left[\right.&\left.E^{r\prime}K_r-\bar{E}^\varphi\bar{K}'_\varphi\right], \nonumber
\end{align}
with $\Gamma_\varphi=-(E^r)^\prime\sqrt{\left|\cos(\delta K_\varphi)\right|}/2\bar{E}^\varphi$ and $K_\varphi=K_\varphi(\bar{K}_\varphi)$. Note that in (\ref{phiT}), the expression for $\bar{K}_\varphi$ cannot be explicitly inverted to give $K_\varphi$ in terms of $\bar{K}_\varphi$. This will not constitute a mayor obstacle for the following calculations. Utilizing the transformed variables, the Poisson bracket between effective Hamiltonian constraints reads 

\begin{equation}
\{H_{eff}[N],H_{eff}[M]\}=C_{eff}\left[\frac{s_cE^r}{(\bar{E}^\varphi)^2}\left(MN'-NM'\right)\right], 
\end{equation}
where $s_c=\sgn\left[\cos(\delta K_\varphi)\right]$, which appears in the calculations since $\left|\cos(\delta K_\varphi)\right|=s_c\cos(\delta K_\varphi)$. The presence of $s_c$ implies a change of signature in the spacetime metric when $s_c=-1$ (see for instance \cite{Asier2}). Looking at the structure function of the past bracket, one is led to propose $q^{11}=s_cE^r/(\bar{E}^\varphi)^2$ for the inverse spatial metric coefficient. Therefore,

\begin{equation}
\bar{ds}^2_{(3)}=\bar{g}_{\mu\nu}^{(3)}dx^\mu dx^\nu=-N^2dt^2+s_c\frac{(\bar{E}^\varphi)^2}{E^r}\left(dr+N_rdt\right)^2+E^rd\Omega^2,
\label{ds2hol}
\end{equation} 
where the change of signature due to $s_c$ is evident.

We proceed next to compute the effective Einstein tensor $\mathscr{G}^{\mu\nu}$. For this case, because of the applied canonical transformation, it is convenient to consider the metric variables $\vec{\Gamma}_P=(N,N_r,E^r,\bar{E}^\varphi)$. The matrix $\mathbf{g}_\Gamma$ is similar in structure to that of (\ref{Syg}), but slightly differs due to the spacetime signature,

\begin{eqnarray}		
		\mathbf{g}_\Gamma=\begin{bmatrix}
		-2N & 0 & 0 & 0 \\
	2s_cN_r(\bar{E}^\varphi)^2/E^r & 2s_c(\bar{E}^\varphi)^2/E^r & 0 & 0 \\
	-s_c(N_r\bar{E}^\varphi/E^r)^2 & -2s_cN_r(\bar{E}^\varphi/E^r)^2 & -s_c(\bar{E}^\varphi/E^r)^2 & 2	\\
	2s_cN_r^2\bar{E}^\varphi/E^r & 4s_cN_rE^\varphi/E^r & 2s_c\bar{E}^\varphi/E^r & 0
\end{bmatrix}.
\nonumber
\end{eqnarray}
On the other hand, the vector	$\vec{S}_\Gamma$ is readily obtained by changing $E^\varphi\rightarrow\bar{E}^\varphi$, $K_\varphi\rightarrow\bar{K}_\varphi$, $\mathcal{H}_c\rightarrow\mathcal{H}_{eff}$ and $\mathcal{C}_c\rightarrow\mathcal{C}_{eff}$ in (\ref{Syg}). The change of signature naturally also affects some of the components of $\mathscr{G}^{\mu\nu}$, in particular,

\begin{eqnarray}
  \mathscr{G}^{01}&=&\frac{-1}{N\bar{E}^\varphi\sqrt{E^r}}\left(\frac{s_cE^r\mathcal{C}_{eff}}{\gamma(\bar{E}^\varphi)^2}-\frac{N_r\mathcal{H}_c}{2N}\right), \nonumber \\
		\mathscr{G}^{11}&=&\frac{s_c\sqrt{E^r}}{2N(\bar{E}^\varphi)^2}\left(\{K_\varphi,H_T^{eff}\}-\dot{K}_\varphi\right)-\frac{N_r}{N\bar{E}^\varphi\sqrt{E^r}}\left(\frac{N_r\mathcal{H}_{eff}}{2N}-\frac{2s_cE^r\mathcal{C}_{eff}}{\gamma(\bar{E}^\varphi)^2}\right). \nonumber
\end{eqnarray}
The remaining components suffer no effect and can be found by doing the same past replacements $E^\varphi\rightarrow\bar{E}^\varphi$, $K_\varphi\rightarrow\bar{K}_\varphi$, $\mathcal{H}_c\rightarrow\mathcal{H}_{eff}$ and $\mathcal{C}_c\rightarrow\mathcal{C}_{eff}$ in (\ref{GEsf}). Evidently the effective constraints (\ref{HDHol}) are also meant to be inserted in $\mathcal{H}_{eff}$ and $\mathcal{C}_{eff}$. For this case, the divergence of the effective Einstein tensor can be written as 

\begin{equation}
\nabla_\mu\mathscr{G}^{\mu\nu}=\sum_{i=0}^2\left[\widetilde{A}_{(i)}^\nu\partial^i_r\left(\dot{E}^r-\{E^r,H_T^{eff}[N,N_r]\}\right)+\widetilde{B}_{(i)}^\nu\partial^i_r\left(\dot{\bar{E}}^\varphi-\{\bar{E}^\varphi,H_T^{eff}[N,N_r]\}\right)\right]+Fn^\nu,
\label{DG3}
\end{equation}
where the explicit expression of $F=F(\delta,\mathbf{\Gamma}_+)$ is quite long and $n^\mu=(t^\mu-N^\mu)/N$ is the unit normal to the spatial hypersurfaces of the foliation\footnote{We are using coordinates such that $t^\mu=(\partial/\partial t)^\mu$.}. Unfortunately, $F$ is such that it does not vanish in solutions of $\dot{E}^r$ and $\dot{\bar{E}}^\varphi$. Despite this, it has a correct classical limit, i.e., $\lim_{\delta\rightarrow0}F(\delta,\mathbf{\Gamma}_+)=0$, or expanding in powers of $\delta$, $F=\mathcal{O}\left(\delta^2\right)$. As in the previous cases, the equations for $\widetilde{A}_{(i)}^\mu$ and $\widetilde{B}_{(i)}^\mu$ are explicitly shown in the final appendix. Due to the additional term containing $F$ we are left with a non-vanishing quantity, 

\begin{equation}
\nabla_\mu\mathscr{G}^{\mu\nu}|_K=Fn^\nu.
\label{DGuv2}
\end{equation}
This means that the theory is not generally covariant with respect to $\bar{g}_{\mu\nu}^{(3)}$, even though the line element was adapted in such a way that the structure of the classical algebra of constraints is displayed by their effective counterparts too.

From (\ref{DGuv2}) one can conclude, nonetheless, that the symmetry regarding spatial diffeomorphisms is not removed. It is clear that for any spatial vector $k^\mu$, we have that $n^\mu k_\mu=0$ and thus, $k_\nu\nabla_\mu\mathscr{G}^{\mu\nu}|_K=0$. The vector $\nabla_\mu\mathscr{G}^{\mu\nu}|_K$ is aligned to the normal and consequently vanishes along any purely spatial direction, i.e., the action is invariant under spatial diffeomorphisms. This of course comes as no surprise since the diffeomorphism constraint is known to generate this type of transformations. The same holds for expression (\ref{DG1}), which corresponds to the model with inverse triad corrections without changes to the classical metric. These particular results do not reveal anything new, but are meant to be seen as consistency requirements of the scheme developed here.

We are going to analyze one final alternative in order to regain general covariance. Hereafter we will restore the phase space dependency of the holonomy parameter $\delta=\delta(E^r)$, and therefore $\gamma_{1,2}=\gamma_{1,2}(K_\varphi,E^r)$. In fact, unlike the precedent holonomy calculations, we will not fix a specific form of the corrections functions, imposing nevertheless only the anomaly-free condition, $$\gamma_2-\gamma_1\frac{\partial\gamma_1}{\partial K_\varphi}+2E^r\frac{\partial\gamma_2}{\partial E^r}=0.$$  As in the inverse triad correction case, the proposal consists in correcting the lapse function appearing in the line element, this is, $N\rightarrow\bar{N}=\sqrt{\left|\partial\gamma_2/\partial K_\varphi\right|}N$. Again, the extra factor multiplying $N$ comes from the deformed structure function in (\ref{HHhol}). We have thus that,

\begin{equation}
\bar{ds}^2_{(4)}=\bar{g}_{\mu\nu}^{(4)}dx^\mu dx^\nu=-s_\gamma\bar{N}^2dt^2+\frac{(E^\varphi)^2}{E^r}\left(dr+N_rdt\right)^2+E^rd\Omega^2,
\label{ds2Esf4}
\end{equation}
with $s_\gamma=\sgn[\partial\gamma_2/\partial K_\varphi]$, hence accounting for a possible change of signature as in $\bar{g}_{\mu\nu}^{(3)}$. This emergent line element has already been considered in \cite{brahma} for the same purpose as here but with a constant holonomy parameter $\delta$. The calculations to find $\nabla_\mu\mathscr{G}^{\mu\nu}$ do not vary significantly from the previous ones. However, it is advantageous to return to the original pair of canonical variables, $K_\varphi$ and $E^\varphi$, so that the space of metric variables can be chosen as $\vec{\Gamma}_P=(\bar{N},N_r,E^r,E^\varphi)$. Therefore 

\begin{eqnarray}
\vec{S}_\Gamma&=&\left[-\frac{\mathcal{H}_{eff}}{\sqrt{\left|\partial\gamma_2/\partial K_\varphi\right|}},-\mathcal{C}_{eff},\dot{K}_r-\{K_r,H_T^{eff}[N,N_r]\},\dot{K}_\varphi-\{K_\varphi,H_T^{eff}[N,N_r]\}\right], \nonumber \\
\mathbf{g}_\Gamma&=&
		\begin{bmatrix}
		-2s_\gamma\bar{N} & 0 & 0 & 0 \\
	2N_r(E^\varphi)^2/E^r & 2(E^\varphi)^2/E^r & 0 & 0 \\
	-(N_rE^\varphi/E^r)^2-2s_\gamma\bar{N}\partial\bar{N}/\partial E^r & -2N_r(E^\varphi/E^r)^2 & -(E^\varphi/E^r)^2 & 2	\\
	2N_r^2E^\varphi/E^r & 4N_rE^\varphi/E^r & 2E^\varphi/E^r & 0
\end{bmatrix},
\end{eqnarray}
All of the components of the modified Einstein tensor $\mathscr{G}^{\mu\nu}$ can be obtained from (\ref{GEsf}) by making the changes $N\rightarrow\bar{N}$, $\mathcal{H}_c\rightarrow s_\gamma\mathcal{H}_{eff}/\sqrt{\left|\partial\gamma_2/\partial K_\varphi\right|}$, and $\mathcal{C}_c\rightarrow\mathcal{C}_{eff}$, except for $\mathscr{G}^{22}$. This component acquires an additional term, namely, $$\mathscr{G}^{22}=\frac{1}{2\bar{N}E^\varphi\sqrt{E^r}}\left[\{K_r,H_T^{eff}\}-\dot{K}_r-\mathcal{H}_{eff}\frac{\partial\bar{N}}{\partial E^r}\left(\left|\frac{\partial\gamma_2}{\partial K_\varphi}\right|\right)^{-1/2}\right]+\frac{1}{4\bar{N}(E^r)^{3/2}}\left(\{K_\varphi,H_T^{eff}\}-\dot{K}_\varphi\right).$$

Interestingly enough, a term featuring the effective Hamiltonian constraint appears in 

\begin{eqnarray}
\nabla_\mu\mathscr{G}^{\mu\nu}&=&\left(\left|\frac{\partial\gamma_2}{\partial K_\varphi}\right|\right)^{-3/2}\sum_{i=0}^2\left[A_{(i)}^\nu\partial^i_r\left(\dot{E}^r-\{E^r,H_T^{eff}[N,N_r]\}\right)+B_{(i)}^\nu\partial^i_r\left(\dot{E}^\varphi-\{E^\varphi,H_T^{eff}[N,N_r]\}\right)\right] \nonumber \\
&&+K^\nu\mathcal{H}_{eff}.
\label{DG4}
\end{eqnarray}
Once more the explicit form of $K^\mu=K^\mu(\mathbf{\Gamma}_+)$ is not too relevant here, but can be consulted in appendix A. We see that now, not only must the Hamilton equations related to $\dot{E}^r$ and $\dot{E}^\varphi$ have to hold, but also one has to restrict the analysis to the constraint surface so that the divergence $\nabla_\mu\mathscr{G}^{\mu\nu}$ vanishes. Note that the on-shell requirement was not needed in the previous cases. Thus, according to the present results, the effective holonomy model is diffeomorphism invariant with respect to the metric $\bar{g}_{\mu\nu}^{(4)}$, but not with respect to $\bar{g}_{\mu\nu}^{(3)}$. Moreover, the lapse function modifications of $\bar{g}_{\mu\nu}^{(4)}$ allowed for covariant and phase space dependent holonomy effects, this is not possible even in the simpler case of a constant $\delta$ parameter if considering a corrected $\bar{E}^\varphi$ variable in $\bar{g}_{\mu\nu}^{(3)}$. The past result can be seen as a generalization of the covariant holonomy model introduced in \cite{brahma}, wherein constant $\delta$ corrections were added to the lapse function. A different method, though, was employed to arrive at this conclusion (see section VI. for additional discussion).

\section{V. Spherical Symmetry and Kantowski-Sachs. A Connection Between the Effective Models}

In the past section we were able to explicitly prove general covariance in quantum corrected spherically symmetric spacetimes. Our objective will now be to find a correspondence between those with inverse triad corrections and the adapted Kantowski-Sachs (KS) model that can describe the exterior region of spherical black holes. In both cases, effective quantum corrections will be taken into account. We start by briefly describing the classical KS spacetime.

\subsection{V.A. Kantowski-Sachs and the Exterior Model}

The KS metric corresponds to anisotropic, but homogenous, cosmological models. It has also been used for the loop quantization of the interior Schwarzschild spacetime \cite{SchwarzschildSingularity,Revisited}. The classical phase space is constituted by two pairs of canonical variables $\mathbf{\Gamma}=\left(b(t),c(t);p_b(t),p_c(t)\right)$. The configuration variables $b$ and $c$ are components of the Ashtekar connection, while their conjugate momenta $p_b$ and $p_c$ respectively, are components of the densitized triad. The line element takes the following form, 

\begin{equation}
ds^2=-N^2dt^2+\frac{p_b^2}{p_c}dr^2+p_cd\Omega^2.
\label{ds2KS}
\end{equation}
The total Hamiltonian consists only of the Hamiltonian constraint, this is,

\begin{equation}
H[N]=\frac{2N}{\gamma^2\sqrt{p_c}}\left[(\gamma^2+b^2)p_b+2bcp_c\right].
\end{equation}
The Poisson brackets for the phase space variables are $$\{b,p_b\}=\gamma, \quad \{c,p_c\}=2\gamma.$$ It is straightforward to verify that $\{H[N],H[M]\}=0$, which in this case, due to the fact that there is no diffeomorphism constraint, reflects the classical algebra of constraints.

Recently, it has been proposed that Wick rotations can be performed in the canonical variables of KS so that the resulting model describes the exterior of the Schwarzschild black hole. The techniques of LQG are then applied to this exterior model following the known procedure for the interior region \cite{AshtekarOlmedoSingh1, AshtekarOlmedoSingh2}. The advantage of this scheme is that one then only has to analyze a Hamiltonian akin to a mechanical model, instead of an inhomogeneous field theory. The rotations are done by setting, $$N=in, \quad p_b=ip_B, \quad b=-iB,$$ and then working in the phase space $\mathbf{\Gamma}=\left(B(t),c(t);p_B(t),p_c(t)\right)$. This can also be seen as an analytic continuation of the previous variables. The line element (\ref{ds2KS}) turns into

\begin{equation}
ds^2=-\frac{p_B^2}{p_c}dr^2+n^2dt^2+p_cd\Omega^2.
\label{ds2Ext}
\end{equation}
Note that now the vector field $\partial/\partial t$ is spacelike, while $\partial/\partial r$ is timelike. Finally, the Hamiltonian changes to

\begin{equation}
H[n]=-\frac{2n}{\gamma^2\sqrt{p_c}}\left[(\gamma^2-B^2)p_B-2Bcp_c\right],
\end{equation}
while the Poisson brackets of the new variables maintain the original structure, $$\{B,p_B\}=\gamma, \quad \{c,p_c\}=2\gamma.$$ Again, it can be easily seen that $\{H[n],H[m]\}=0$. The consequence of implementing the Wick rotations is simply to shift from a foliation with spacelike hypersurfaces, to one with timelike hypersurfaces. Hereafter, we will refer to this prescription as the exterior KS model.

\subsection{V.B. Quantum Corrections in the Exterior KS Model}

Having introduced the properties of the classical exterior KS spacetime, we will be interested as well in an effective Hamiltonian that incorporates quantum effects into its dynamics. In fact, we shall first consider the most general effective Hamiltonian consistent with homogeneity (no spatial dependence on the phase space variables), this is,

\begin{equation}
H_{eff}[n]=nF(B,c;p_B,p_c).
\label{HeffKS}
\end{equation}
If one tests general covariance with the method described in this paper, it turns out that the effective Hamiltonian (\ref{HeffKS}) describes a model whose action is diffeomorphism invariant with respect to the classical metric (\ref{ds2Ext}). The calculations go along the same lines as the previously done analysis. For the sake of brevity and since the procedure has been illustrated numerous times in this work, we just show here the end result but comment on the initial steps. Due to the absence of a diffeomorphism constraint, the metric of interest has now three independent coefficients. The same applies for the Einstein tensor as well. All of the mathematical objects of (\ref{Syg}) will then be of dimension $n=3$. In particular, the space of metric variables is chosen as $\vec{\Gamma}_P=(n,p_B,p_c)$ and $\vec{S}_\Gamma=(-F,\dot{B}-\{B,H_{eff}[n]\},\dot{c}-\{c,H_{eff}[n]\})$, where a dot denotes derivation with respect to the coordinate $t$. The rest follows from equation (\ref{SG}). It can be found thus that, $$\nabla_\mu\mathscr{G}^{\mu\nu}=\frac{1}{2n^3p_B\sqrt{p_c}}\left[\dot{c}\left(\dot{p}_c-\{p_c,H_{eff}[n]\}\right)+\dot{B}\left(\dot{p}_B-\{p_B,H_{eff}[n]\}\right)\right]\delta_0^\nu.$$ It is clear that this quantity vanishes for solutions of the pair of Hamilton equations, $\dot{p}_B=\{p_B,H\}$ and $\dot{p}_c=\{p_c,H\}$, i.e., $\nabla_\mu\mathscr{G}^{\mu\nu}|_K=0$ keeping the notation introduced in previous sections. Covariance is therefore a guaranteed feature of effective models even as general as (\ref{HeffKS}). This property will be exploited from this point forward. From the perspective of the algebra of effective constraints, it is trivial to realize that $\{H_{eff}[n],H_{eff}[m]\}=0$, and so, the structure of the classical algebra is unmodified.

\subsection{V.C. Connecting the Effective Dynamics of Inverse Triad Corrections}

So far we have seen that (i) the effective theories given by (\ref{HeffKS}) along with the exterior KS metric, and (ii) the spherically symmetric Hamiltonian corrected by inverse triad effects with an emergent metric, are both invariant under diffeomorphisms. They also can describe the same physical objects and situations. Is there a way to relate one to another and make the two equivalent? We will show that this is the case by proving that their dynamics are equal and yield the same line element. Particularly, we shall seek the exterior KS modified model that corresponds to the two metrics studied in section IV.B., i.e, $\bar{g}_{\mu\nu}^{(1)}$and $\bar{g}_{\mu\nu}^{(2)}$. To this end, let us consider the following effective Hamiltonian that contains inverse triad corrections,

\begin{equation}
H_{eff}[n]=-\frac{2n}{\gamma^2\sqrt{p_c}}\left[(\gamma^2-B^2)p_B\beta_1(p_c)-2Bcp_c\beta_2(p_c)\right].
\label{HeffKSinv}
\end{equation}
In a similar sense as in (\ref{HeffInv}), the presence of the functions $\beta_{1,2}(p_c)$ is supposed to account for the inverse triad effects implied by LQG. Analogously to the case of spherical symmetry, this Hamiltonian is obtained by replacing $$\frac{1}{\sqrt{p_c}}\rightarrow\frac{\beta_1(p_c)}{\sqrt{p_c}} \quad \text{and} \quad \sqrt{p_c}\rightarrow\beta_2(p_c)\sqrt{p_c}$$ in the classical exterior KS model. In fact, we are going to compare the equations of motion yielded by the mentioned spherical case to those found with the effective Hamiltonian (\ref{HeffKSinv}), these are

\begin{equation}
\dot{p}_B=\frac{-n}{\gamma\sqrt{p_c}}\left(\beta_1Bp_B+\beta_2cp_c\right), \quad \dot{p}_c=-\frac{2}{\gamma}n\beta_2B\sqrt{p_c}, \quad \dot{B}=\frac{n\beta_1(B^2-\gamma^2)}{2\gamma\sqrt{p_c}}, \quad (\gamma^2-B^2)p_B\beta_1-2Bcp_c\beta_2=0.
\label{EoMExt}
\end{equation}
The first two expressions provide the components of the Ashtekar connection, $B$ and $c$, in terms of the densitized triad and its time derivatives, namely,

\begin{equation}
B=-\frac{\gamma\dot{p}_c}{2n\beta_2\sqrt{p_c}}, \quad c=\frac{\gamma}{\beta_2np_c}\left(\frac{\beta_1p_B\dot{p}_c}{2\beta_2\sqrt{p_c}}-\dot{p}_B\sqrt{p_c}\right).
\end{equation}

Let us begin with the spherically symmetric effective line element $\bar{ds}^2_{(1)}$, recall that it was yielded by changing $E^r$ to $\bar{E}^r=\alpha_2^2E^r$ in the classical metric. A natural exterior KS geometry to search for the desired correspondence is then 

\begin{equation}
\bar{ds}^2_{(5)}=\bar{g}_{\mu\nu}^{(5)}dx^\mu dx^\nu=-\frac{p_B^2}{\bar{p}_c}dr^2+n^2dt^2+\bar{p}_cd\Omega^2,
\label{ds2Ext2}
\end{equation} 
where $\bar{p}_c=\beta_2^2p_c$. The analogy between this equation and (\ref{ds2Esf2}) is immediate. Choosing a gauge in which $p_c=t^2$, and inserting the previous expressions of the Ashtekar connection into the last two equations of (\ref{EoMExt}), we obtain 

\begin{equation}
   n^3+\frac{1}{\beta_2^2}\left(\frac{4t^2}{\beta_1}\frac{d\beta_2}{dp_c}-1\right)n+\frac{2t}{\beta_1\beta_2}\dot{n}=0, \quad \left(2-\frac{\beta_1}{\beta_2}\left[1+\beta_2^2n^2\right]\right)\frac{p_B}{\beta_2t}+2t\frac{d}{dt}\left(\frac{p_B}{\beta_2t}\right)+\frac{4t^2}{\beta_2}\frac{d\beta_2}{dp_c}\left(\frac{p_B}{\beta_2t}\right)=0,
	\label{EoMExt2}
\end{equation}
Once this pair of equations is solved, the effective exterior KS metric is completely determined for a given set of explicit correction functions $\beta_{1,2}(p_c)$. Nevertheless, we will not need to go as far as that since the dynamics of the theory is already contained in these equations of motion.

We now return to the spherically symmetric model with inverse triad corrections treated in subsection IV.B. Let us examine the equations of motion found from the effective Hamiltonian (\ref{HeffInv}) and compare their form to those of its exterior KS counterpart. We have already solved two of them since they were needed in our general covariance analysis, see (\ref{K2}). The remaining equations of motion we require for this purpose read,

\begin{align}
(E^r)'K_r-E^\varphi K'_\varphi=0, \quad \frac{K_\varphi}{2\gamma^2}\left(4\alpha_2K_rE^r+\alpha_1K_\varphi E^\varphi\right)+2\alpha_1E^\varphi(1-\Gamma_\varphi^2)+4\alpha_2E^r\Gamma'_\varphi=0, \nonumber \\
\dot{K}_\varphi=\frac{1}{2\gamma^2(E^\varphi)^2\sqrt{E^r}}\left[4\gamma N_rK'_\varphi(E^\varphi)^2\sqrt{E^r}+2\gamma^2(E^r)'\left(N\left[(E^r)^2\right]'\frac{d\alpha_2}{dE^r}+\alpha_2\left[(NE^r)'+N'E^r\right]\right)\right.& \nonumber \\
\left.-N\alpha_1\left([4\gamma^2+K_\varphi^2][E^\varphi]^2+\left[\gamma(E^r)'\right]^2\right)\right],
\end{align}
where the first and second expressions correspond to the diffeomorphism constraint and the Hamiltonian constraint, respectively. We are looking for static solutions, therefore, to solve the equations of motion the gauge $N_r=0$ can be fixed and the $t$ dependence of the canonical variables can be dropped. This implies that $K_\varphi=K_r=0$ and consequently, the diffeomorphism constraint is trivially satisfied. Implementing the areal gauge $E^r=r^2$, the past expressions can be rearranged as,

\begin{align}
   \left(\frac{E^\varphi}{\alpha_2r}\right)^3+\frac{1}{\alpha_2^2}\left(\frac{4r^2}{\alpha_1}\frac{d\alpha_2}{dE^r}-1\right)\frac{E^\varphi}{\alpha_2r}+\frac{2r}{\alpha_1\alpha_2}\left(\frac{E^\varphi}{\alpha_2r}\right)'&=0, \nonumber \\ 
	\left(2-\frac{\alpha_1}{\alpha_2}\left[1+\alpha_2^2\left(\frac{E^\varphi}{\alpha_2r}\right)^2\right]\right)N+2rN'+\frac{4r^2}{\alpha_2}\frac{d\alpha_2}{dE^r}N&=0.
	\label{EoMEsf}
\end{align}
If the time coordinate of the exterior KS metric is identified with the radial coordinate of the spherically symmetric model ($r=t$, $E^r=p_c$), which can be done due to both of the canonical vector fields $\partial/\partial t$ and $\partial/\partial r$ being spacelike, the dynamics of each theory can be compared. Indeed, the comparison between the line elements $\bar{ds}^2_{(5)}$ in (\ref{ds2Ext2}) and $\bar{ds}^2_{(1)}$ in (\ref{ds2Esf3}) is immediately direct. The two set of equations of motion, (\ref{EoMExt2}) and (\ref{EoMEsf}), yield the same metric, so long as 

\begin{equation}
\alpha_1=\beta_1 \quad \text{and} \quad \alpha_2=\beta_2.
\label{alphabeta1} 
\end{equation}
Thus, the corresponding exterior KS model of an effective spherically symmetric model, characterized by corrections $\alpha_{1,2}$, is that whose own modifications $\beta_{1,2}$ satisfy $\alpha_1=\beta_1$ and $\alpha_2=\beta_2$. Evidently, the converse is also possible, i.e., determining effective spherical symmetry by starting with a modified exterior KS model making sure that (\ref{alphabeta1}) holds.

Consider next the following modified exterior KS line element, 

\begin{equation}
\bar{ds}^2_{(6)}=\bar{g}_{\mu\nu}^{(6)}dx^\mu dx^\nu=-\frac{p_B^2}{p_c}dr^2+\bar{n}^2dt^2+p_cd\Omega^2,
\label{ds2Ext3}
\end{equation}
with $\bar{n}=\beta_2n$. As in the last case, one can likewise expect that it should be possible to match it to the similarly modified spherical metric $\bar{g}_{\mu\nu}^{(2)}$ of (\ref{ds2Esf3}), i.e., the one obtained by changing $N$ to $\bar{N}=\alpha_2N$ in the classical description of the theory. Following the previous procedure, the equations of motion for the effective exterior KS model and effective spherical symmetry can be respectively expressed as

\begin{equation}
\bar{n}^3-\bar{n}+\frac{2\beta_2}{\beta_1}t\dot{\bar{n}}=0, \quad \left[2-\frac{\beta_1}{\beta_2}\left(1+\bar{n}^2\right)\right]\frac{p_B}{t}+2t\frac{d}{dt}\left(\frac{p_B}{t}\right)=0,
\end{equation}
and

\begin{equation}
 \left(\frac{E^\varphi}{r}\right)^3-\frac{E^\varphi}{r}+\frac{2\alpha_2}{\alpha_1}r\left(\frac{E^\varphi}{r}\right)'=0, \quad \left[2-\frac{\alpha_1}{\alpha_2}\left(1+\left[\frac{E^\varphi}{r}\right]^2\right)\right]\bar{N}+2r\bar{N}'=0.
\end{equation}
It is clear that if,

\begin{equation}
\frac{\beta_2}{\beta_1}=\frac{\alpha_2}{\alpha_1},
\label{alphabeta}
\end{equation}
then both effective dynamics describe the same line element. We have found, therefore, another exterior KS and spherical symmetry effective correspondence, where the same conclusion as in the first pair of models holds, but with equation (\ref{alphabeta}) being the relation between corrections.

Note that we were able to generally assign a pair of corrections $\beta_{1,2}$ to another pair $\alpha_{1,2}$, due to the use of emergent line elements in which $E^r\rightarrow\bar{E}^r$ and $p_c\rightarrow\bar{p}_c$ for the first case, while $N\rightarrow\bar{N}$ and $n\rightarrow\bar{n}$ for the second one. Otherwise, the correspondence is not possible, for example if one solves the effective dynamics of both models and then tries to match the spherically symmetric classical metric (\ref{ds2Esf}) to the modified exterior line element $\bar{ds}^2_{(5)}$ of (\ref{ds2Ext2}), or viceversa, matching the classical KS metric (\ref{ds2Ext}) to modified spherical symmetry given by $\bar{g}_{\mu\nu}^{(2)}$ in (\ref{ds2Esf3}). There is one final matter to mention, namely, the diffeomorphism invariance of the action implied by (\ref{HeffKSinv}) with respect to the spacetimes described by the recently introduced line elements $\bar{ds}^2_{(5)}$ and $\bar{ds}^2_{(6)}$. We have already proved this property for the classical metric (\ref{ds2Ext}) combined with that same effective model, and also for the inverse triad corrected spherically symmetric geometry. It turns out that the effective exterior KS model is generally covariant when utilizing these two final effective exterior metrics too. The proofs are analogous to the covariance results of section IV.B and the adaptation to the exterior KS inverse triad corrected case is straightforward.

\subsection{V.D. Connecting the Effective Dynamics of Holonomy Corrections}

We may try now to obtain a similar relation between the exterior KS model and spherical symmetry, this time with holonomy modifications. For this purpose let us consider the KS Hamiltonian constraint analogous to (\ref{HeffHol}),

\begin{equation}
H_{eff}[n]=-\frac{2n}{\gamma^2\sqrt{p_c}}\left[\left(\gamma^2-\lambda_1^2(p_c,B)\right)p_B-2\lambda_2(p_c,B)cp_c\right].
\label{HeffKShol}
\end{equation}
If $\lambda_{1,2}=B$, the classical description is recovered. Explicit holonomy correction functions for $\lambda_{1,2}$ can be found in \cite{AshtekarOlmedoSingh1, AshtekarOlmedoSingh2}, and in \cite{Vandersloot} for the corresponding interior model. The general idea of the previous subsection is unchanged: examine the effective dynamics of the two models and find out under what conditions they coincide, i.e., describe the same spacetime metric. Unfortunately the analysis is not as conclusive as the above inverse triad case. The holonomy modified Hamiltonian (\ref{HeffKShol}) yields the following effective field equations

\begin{equation}
\dot{p}_B=\frac{-n}{\gamma\sqrt{p_c}}\left(\lambda_1\frac{\partial\lambda_1}{\partial B}p_B+\frac{\partial\lambda_2}{\partial B}cp_c\right), \quad \dot{p}_c=-\frac{2}{\gamma}n\lambda_2\sqrt{p_c}, \quad \dot{B}=\frac{n(\lambda_1^2-\gamma^2)}{2\gamma\sqrt{p_c}}, \quad (\gamma^2-\lambda_1^2)p_B-2\lambda_2cp_c=0.
\label{EoMExtHol}
\end{equation}
The first obstacle is that, unlike the inverse triad effective dynamics, without the explicit holonomy function $\lambda_2$ we are not able to express $B$ in terms of the rest of the variables. At most, one can get from the first two equations,

\begin{equation}
\lambda_2=-\frac{\gamma\dot{p}_c}{2n\sqrt{p_c}}, \quad c\frac{\partial\lambda_2}{\partial B}=-\frac{\gamma}{np_c}\left(\frac{n\lambda_1}{\gamma}\frac{\partial\lambda_1}{\partial B}p_B+\dot{p}_B\sqrt{p_c}\right).
\end{equation}
In the last subsection it was proven that the two lapse modified line elements of the models of interest obey the same effective dynamics. Moreover, in subsection IV.C. it was shown that the holonomy modified theory is covariant with respect to metric $\bar{g}_{\mu\nu}^{(4)}$, which contains a corrected lapse $\bar{N}$ (see equation (\ref{ds2Esf4})). Following these previous results, it is reasonable to propose an exterior KS metric of the type (\ref{ds2Ext3}) with $\bar{n}=\lambda_3(B,p_c)n$, so that the desired connection to spherical symmetry could be found. We shall let $\lambda_3$ be an arbitrary correction function. Another point to take into account is the change of signature indicated by (\ref{ds2Esf4}) for $\bar{g}_{\mu\nu}^{(4)}$. In order to consistently relate that metric to an exterior KS one, which has signature $-1$, the analysis has to be restricted to corrections functions $\gamma_2$ such that $s_\gamma=\sgn\left[\partial\gamma_2/\partial K_\varphi\right]=1$. With this in mind and fixing the gauge to $p_c=t^2$, we can write the last two expressions in (\ref{EoMExtHol}) as,

\begin{eqnarray}
\left(1-\frac{\lambda_1^2}{\gamma^2}\right)\bar{n}^3-\frac{4t^2}{\gamma\lambda_3}\frac{\partial\lambda_2}{\partial p_c}\bar{n}^2+2t\left(\dot{\bar{n}}-\frac{\dot{\lambda}_3}{\lambda_3}\bar{n}\right)&=&0, \nonumber \\
\frac{\bar{n}^2}{\lambda_3^2}\left[\left(\frac{\lambda_1^2}{\gamma^2}-1\right)\frac{\partial\lambda_2}{\partial B}+2\left(1+\frac{\bar{n}\lambda_1}{\gamma\lambda_3}\frac{\partial\lambda_1}{\partial B}\right)\right]\frac{p_B}{t}+2t\frac{d}{dt}\left(\frac{p_B}{t}\right)&=&0.
\label{EoMExtHol2} 
\end{eqnarray}
Note that in this gauge, $\lambda_2(B,p_c)=-\gamma/n$. The past equations are analogous to the field equations in (\ref{EoMExt2}), with the difference that we would need to insert explicit $\lambda(B,p_c)$ functions and use $\lambda_2(B,p_c)=-\gamma/n$ in order to remove the $B$ dependency. For reference, if $\lambda_1=\lambda_2=B=-\gamma/n$ and $\lambda_3=1$, the classical field equations are obtained.

Let us turn back to the effective spherically symmetric model, particularly, to the one that contains holonomy corrections described by the Hamiltonian (\ref{HeffHol}). Once again, we are interested in static solutions of the Hamilton equations and therefore, we set the gauge to $N_r=0$, $E^r=r^2$, and time independent canonical variables. A straightforward process then reveals that the relevant effective equations of motion are given by

\begin{align}
\gamma_2=&0, \\
\left(1+\frac{\gamma_1^2}{4\gamma^2}\right)\left(\frac{E^\varphi}{r}\right)^3-\frac{E^\varphi}{r}+2r\left(\frac{E^\varphi}{r}\right)'=&0, \quad \left[1-\left(\frac{E^\varphi}{r}\right)^2\left(1+\frac{\gamma_1^2}{4\gamma^2}\right)-2r\frac{\gamma_3'}{\gamma_3}\right]\bar{N}+2r\bar{N}'=0,
\label{EoMEsfHol}
\end{align}
where $\bar{N}=\gamma_3N$ and $\gamma_3=\sqrt{\left|\partial\gamma_2/\partial K_\varphi\right|}$. As in the previous exterior KS effective dynamics, we encounter $K_\varphi$ dependence due to the appearance of the $\gamma_1(K_\varphi,E^r)$ function. Such dependence is removed when solving $\gamma_2(K_\varphi,E^r)=0$ once an explicit form of the correction function is given. In fact, this on-shell condition for $\gamma_2$ can be seen as yielding a $K_\varphi(E^r)$ dependence, which in turn implies that on-shell the other $\gamma$ functions depend at most on $E^r=r^2$. In the classical model $\gamma_1=\gamma_2=K_\varphi=0$ for this gauge. 

At this point it is clear that in order to compare the obtained field equations for both models, namely, the two expressions in (\ref{EoMExtHol2}) and (\ref{EoMEsfHol}) each, we require explicit forms of the correction functions. As a first approach we can consider an example motivated by the exterior KS model proposed in \cite{AshtekarOlmedoSingh1,AshtekarOlmedoSingh2}. The holonomy modifications there are characterized by\footnote{In the mentioned references an additional correction related to the connection variable $c$ is also included. Here, it will not be considered since its corresponding variable in the spherically symmetric model is $K_r$, such kind of modifications of the Hamiltonian are not studied in this paper.} $$\lambda_1=\lambda_2=\frac{1}{\delta}\sinh(\delta B),$$ with the standard exterior KS line element given by (\ref{ds2Ext}), which makes $\lambda_3=1$ for this case. Our purpose is to find functions $\gamma_1$ and $\gamma_2$ such that the effective dynamics of both models are equivalent. Inserting $\lambda_1$ and $\lambda_2$ into the field equations (\ref{EoMExtHol2}) we readily obtain

\begin{equation}
\bar{n}^3-\bar{n}+2t\dot{\bar{n}}=0, \quad \left[2\bar{n}^2+(1-3\bar{n}^2)\sqrt{1+\frac{\gamma^2\delta^2}{\bar{n}^2}}\right]\frac{p_B}{t}+2t\frac{d}{dt}\left(\frac{p_B}{t}\right)=0.
\label{EoMExtHol3}
\end{equation}
 This allows us to make the sought comparison between the last pair of equations and those in (\ref{EoMEsfHol}). Recall that we identify the variables $t\leftrightarrow r$, $\bar{n}\leftrightarrow E^\varphi/r$ and $p_B/t\leftrightarrow\bar{N}$. It can be easily realized that if $\gamma_1=0$, i.e., if the $\gamma_1$ holonomy modification vanishes on-shell, then the equations for $\dot{\bar{n}}$ and $(E^\varphi/r)'$ do coincide. However, even in this simple model, the remaining expressions (those involving $d(p_B/t)/dt$ and $\bar{N}'$) are clearly incompatible since $\gamma_3=\gamma_3(E^r)$ on-shell. The same conclusions holds even if $\lambda_3\neq1$, where long algebraic calculations reveal that there does not exist a function $\lambda_3(B,p_c)$ such that the two equations in (\ref{EoMEsfHol}) and those in (\ref{EoMExtHol3}) can be exactly matched. Therefore, there is not a spherically symmetric model with holonomy corrections described by Hamiltonian (\ref{HeffHol}) that corresponds to an exterior KS model modified by holonomy functions $\lambda_1=\lambda_2=\sinh(\delta B)/\delta$. However it still might possible that other covariant effective models with this kind of symmetry, for instance those of \cite{zhang,belfaqih,Asier2}, might be connected to this particular exterior KS effective theory. In a few words, at this point of the analysis, it cannot be guaranteed that one can always find a covariant spherically symmetric model for any effective KS interior metric that has been extended to the exterior region through an analytic continuation.

\section{VI. Discussion} 

In this work we have focused on the issue of the breakdown of general covariance in effective spherically symmetric models proposed by LQG. First, we introduced a general criterion that can be readily applied to any canonical formulation of a gravitational theory in the vacuum. In fact, while the main objective of the paper is the study of semi-classical models yielded by LQG, the criterion is broad enough as to include any kind of effective theory. It is based on the well-known Bianchi identity $\nabla_\mu G^{\mu\nu}=0$ implied by invariance under diffeomorphisms in GR. In our case, the criterion turns into $$\nabla_\mu\mathscr{G}^{\mu\nu}|_K=0,$$ where $\mathscr{G}^{\mu\nu}$ contains a suitable combination of the Hamilton equations of motion, and $|_K$ denotes evaluation in those solutions that express conjugate momenta in terms of configuration variables.  

The derived criterion was applied to spherically symmetric models with quantum corrections due to inverse triad and holonomy effects. These corrections were assumed to be contained in functions that modify the Hamiltonian of the classical theory. Both cases were treated separately. For inverse triad corrections, general covariance was recovered only after the classical line element itself was also altered by the considered correction functions, this alternative concept was recently named as emergent line element. An unmodified metric leads to the loss of general covariance. This is in whole agreement with previous results, for instance, \cite{Tibrewala_2012,Tibrewala_2014}. Here, two different families of effective metrics were found to be diffeomorphism invariant when an inverse triad modified Hamiltonian is used to describe the semi-classical theory. It is worth highlighting that this is possible without the need of imposing specific expressions for the corrections functions. Likewise, for the case of holonomy modifications, two emergent metrics were explored. In the first one, the spatial metric featured a correction factor and was proven to break covariance. At most, this invariant property was demonstrated only for spatial diffeomorphisms. This provides an interesting example in which the classical structure of the algebra of constraints is maintained, but general covariance is not. In contrast, in the second proposal the lapse of the line element was modified, ultimately leading to invariance under general diffeomorphisms.

Finally, we established a connection between the inverse triad effective versions of spherical symmetry, and a Kantowski-Sachs model that has been transformed through Wick rotations in order to describe spatially inhomogeneous spacetimes (here referred to as exterior KS model). This connection is simply a map between the correction functions of the spherically symmetric model and their counterparts in the exterior KS Hamiltonian. No prior assumptions were made regarding these functions and any inverse triad modification studied here can be mapped to a covariant effective exterior KS model. This seems to be in contradiction with the no-go result found in \cite{Nogo}. However, the incorporation of emergent metrics, which is not considered in the previous reference, is of crucial importance for the recovery of general covariance and is also what enables the correspondence. A similar connection was attempted for the case of holonomy corrections with less general results as those found in the inverse triad analysis. A map between corrections of both theories could not be explicitly built. In fact, considering an exterior KS proposal motivated by \cite{AshtekarOlmedoSingh1,AshtekarOlmedoSingh2}, we found that it is not possible to match it to a covariant spherically symmetric theory, at least one described by the effective models studied in this paper. Of course, other models such as the ones presented in \cite{zhang,belfaqih,Asier2} remain as an option for this purpose.

Another interesting aspect to point out is that, despite the fact that in the exterior KS case, a generally covariant model can be obtained by considering an effective Hamiltonian, solving the effective dynamics, and then constructing the metric following its classical expression, it is also possible to yield an equally covariant model by doing that same procedure, but modifying the metric itself in addition. Not only that, but according to the results found here, the modified metric can be matched to a likewise effective spherically symmetric model, while the unmodified metric cannot, even taking into account the effective dynamics. One can again, perform Wick rotations to this KS spacetime to return to a cosmological model. This can hint toward, for instance, effective cosmological models that may be different to those already available in the literature when taking into account emergent forms of the metric to describe them. For future work it may be worthy to explore the properties of this type of spacetimes and whether they possess the same qualitative features of their predecessors. The most important attribute of the models with modified line elements here being their slicing independence, a defining characteristic of the classical theory.

\subsection{VI.A. Comparison with Similar Works}  

In essence, our work is consistent with the other covariance approaches that have been reported so far in the literature. In the following paragraphs we provide a more detailed comparison. The framework developed here can also be seen as a complement to those of \cite{zhang,brahma,belfaqih,Asier2} since the modified Einstein tensor $\mathscr{G}_{\mu\nu}$, which replaces the classical $G_{\mu\nu}$ in our approach, contains information about the effective metric with corrections. Moreover the constraint algebra is required to hold too. The covariance perspective taken in those references is that of verifying that the gauge transformations of the metric, generated by the modified constraints, correspond to diffeomorphisms, while we explore covariance through the diffeomorphism invariance of the action of the effective theory. Admittedly, even though the calculations required for our scheme are more intricate compared to those of \cite{zhang,zhang2,brahma,belfaqih}, it proves general covariance in a certainly explicit manner through the vanishing of the divergence of the modified field equations. The relation between invariance under diffeomorphisms of the action is direct.

As far as the effective models examined in this paper, both perspectives yield the same results concerning the covariance of the theory. In \cite{zhang} for example, covariance can be regained if one modifies the spatial metric by the correction factor that appears in the algebra. The mentioned factor can depend on certain combination of phase space variables, two of which are the densitized triad component $E^r$ and the configuration variable $K_\varphi$. However, sole dependency of the correction factor on $K_\varphi$ is not compatible with the covariance conditions stated there, while sole dependency of that same factor on $E^r$ is. This is in agreement with our inverse triad results because, in the analysis made here, the correction functions only depended on precisely the $E^r$ component. On the other hand, the model featuring holonomy corrections that depend only on the extrinsic curvature component $K_\varphi$ is not diffeomorphism invariant with respect to the metric with corrections in the $q_{11}$ radial component. Covariance was recovered upon modifying the component involving the lapse instead. This alternative modification is not studied in \cite{zhang}, reference \cite{brahma} has done it with the same conclusion but different method. In this regard, we have extended that result to holonomy corrections with a phase space dependent parameter. Returning to reference \cite{zhang}, it is worth highlighting that applying our scheme to the first effective model studied there, it can be checked that it is indeed diffeomorphism invariant. A conclusion which is consistent with their results. The general case presented there, though, cannot be studied yet with the treatment described in this paper due to different considerations on the dependence of the correction factor appearing in the metric. In particular, the possibility that said factor could depend on radial derivatives of phase space variables is not considered here. Adapting our framework to include such dependence could be of future interest. As for \cite{Deformed1,Deformed2} the relation with our work seems to lie also in the use of the constraint algebra to restrict consistent deformations. Further work is required to analyze other specific models within spherical symmetry, as well as other inhomogeneous models like those of Gowdy.

When comparing the two discussed covariance approaches one may, however, notice some basic differences in structure. For instance, our criterion can involve as high as third order derivatives of the metric due to the computation of $\nabla_\mu\mathscr{G}^{\mu\nu}$, while the diffeomorphisms generated by constrictions condition contains at most second order derivatives. It remains to be seen if they could disagree for a particular effective action, as well as establishing in a mathematically precise way if both are equivalent or if one is implied by the other. Such a connection is worth pursuing and is left for future work.

\section{Appendix. The Phase Space Coefficients of $\nabla_\mu\mathscr{G}^{\mu\nu}$}
\renewcommand{\theequation}{A.\arabic{equation}}
\setcounter{equation}{0}

In this appendix we explicitly show the coefficients that appear in the right hand side of the divergence of the modified spherically symmetric Einstein tensor. This is done for all the cases studied in the main text. While some expressions might be lengthy, we include them as they could be of future interest in the structure of the covariance problem within effective models, as well as to give the reader a precise view of their content. For the sake of clarity, we will repeat the relevant equations of the previous sections.

\subsection{A. Classical Description}

Let us first define the following quantities,

\begin{eqnarray}
A_{(0)}&=&\frac{1}{2\gamma N^2E^r(E^\varphi)^3}\left[\gamma E^\varphi\left(N'(E^r)'+2N''E^r\right)-2\gamma N'E^r(E^\varphi)'+2\sqrt{\left|E^r\right|}(E^\varphi)^2\left((N_rK_r)'-\dot{K}_r\right)\right], \nonumber \\
B_{(0)}&=&\frac{1}{2\gamma N^2E^r(E^\varphi)^4}\left[\gamma E^\varphi\left[2\left(NE^r(E^r)'\right)'-N(E^r)'^2\right]-4\gamma NE^r(E^r)'(E^\varphi)'+2\sqrt{\left|E^r\right|}(E^\varphi)^3\left(N_rK_\varphi'-\dot{K}_\varphi\right)\right], \nonumber \\
A_{(1)}&=&\frac{1}{2N(E^\varphi)^3E^r}\left[2E^r\left(E^\varphi\right)'-\left(E^r\right)'E^\varphi\right].
\label{ACl}
\end{eqnarray}
It will also be convenient to utilize the purely spatial vectors, which are written in the $\{t,r,\theta,\varphi\}$ basis,

\begin{equation}
l_{(1)}^\mu=\left[0,\frac{K_\varphi'\sqrt{\left|E^r\right|}}{\gamma N\left(E^\varphi\right)^3},0,0\right], \quad l_{(2)}^\mu=\left[0,-\frac{K_r\sqrt{\left|E^r\right|}}{\gamma N\left(E^\varphi\right)^3},0,0\right].
\label{lmu}
\end{equation}

In the classical analysis of the Einstein tensor of subsection IV.A we expressed its divergence in equation (\ref{DGuv}), this is,

\begin{equation*}
\nabla_\mu\mathscr{G}^{\mu\nu}=\sum_{i=0}^2\left[A_{(i)}^\nu\partial^i_r\left(\dot{E}^r-\{E^r,H_T[N,N_r]\}\right)+B_{(i)}^\nu\partial^i_r\left(\dot{E}^\varphi-\{E^\varphi,H_T[N,N_r]\}\right)\right].
\tag{\ref{DGuv}}
\end{equation*}
The vectors $A_{(i)}^\mu$ and $B_{(i)}^\mu$ are then given by,

\begin{align}
A_{(0)}^\mu=&A_{(0)}n^\mu, \quad B_{(0)}^\mu=B_{(0)}n^\mu+l_{(1)}^\mu, \quad A_{(1)}^\mu=A_{(1)}n^\mu+l_{(2)}^\mu, \nonumber \\
B_{(1)}^\mu=&\frac{\left(E^r\right)'}{N\left(E^\varphi\right)^3}n^\mu, \quad A_{(2)}^\mu=-\frac{1}{\left(E^\varphi\right)^2}n^\mu, \quad B_{(2)}^\mu=0, \nonumber
\end{align}
where $n^\mu=(t^\mu-N^\mu)/N$ is the unit normal to the spatial hypersurfaces of the foliation.

\subsection{B. Inverse Triad Corrections} 

Similarly, let us introduce the auxiliary quantities,

\begin{eqnarray}
\bar{A}_{(0)}&=&\frac{1}{2\gamma N^2E^r(E^\varphi)^3}\left[\gamma E^\varphi\left(N'(E^r)'(2\alpha_2-\alpha_1)+2N''E^r\alpha_2\right)-2\gamma N'E^r(E^\varphi)'\alpha_2+2\sqrt{\left|E^r\right|}(E^\varphi)^2\left((N_rK_r)'-\dot{K}_r\right)\right. \nonumber \\
&&\hspace{2.5cm}\left.+\gamma(E^\varphi)'\left(N(E^r)'(\alpha_1-\alpha_2)-2NE^r(E^r)'\frac{d\alpha_2}{dE^r}\right)+\gamma N(E^r)'^2E^\varphi\left(2E^r\frac{d^2\alpha_2}{d(E^r)^2}-\frac{d\alpha_1}{dE^r}\right)\right. \nonumber \\
&&\hspace{2.5cm}\left.+\gamma NE^\varphi\left(\frac{(E^r)'^2}{2E^r}-(E^r)''\right)\left(\alpha_1-\alpha_2+4E^r\frac{d\alpha_2}{dE^r}\right)+4\gamma N'E^r(E^r)'E^\varphi\frac{d\alpha_2}{dE^r}\right], \nonumber \\
\bar{B}_{(0)}&=&\frac{1}{2\gamma N^2E^r(E^\varphi)^4}\left[\gamma E^\varphi\left[2\left(NE^r(E^r)'\right)'-N(E^r)'^2\right]\alpha_2-4\gamma NE^r(E^r)'(E^\varphi)'\alpha_2+2\sqrt{\left|E^r\right|}(E^\varphi)^3\left(N_rK_\varphi'-\dot{K}_\varphi\right)\right. \nonumber \\
&&\hspace{2.5cm}\left.+2\gamma NE^r(E^r)'^2E^\varphi\frac{d\alpha_2}{dE^r}\right], \nonumber \\
\bar{A}_{(1)}&=&\frac{1}{2N(E^\varphi)^3E^r}\left[2\alpha_2E^r\left(E^\varphi\right)'-\alpha_1\left(E^r\right)'E^\varphi\right].
\end{eqnarray}
Notice that if $\alpha_1=\alpha_2=1$, we recover the classical set of equations (\ref{ACl}). In subsection IV.B. we studied different metric proposals with the objective of restoring covariance in the inverse triad effective model. We present the explicit results for those cases.

Using the classical line element (\ref{ds2Esf}) we showed the form of the divergence of the modified Einstein tensor in equation (\ref{DG1}), this is, 

\begin{align*}
\nabla_\mu\mathscr{G}^{\mu\nu}=&\sum_{i=0}^2\left[\bar{A}_{(i)}^\nu\partial^i_r\left(\dot{E}^r-\{E^r,H_T^{eff}[N,N_r]\}\right)+\bar{B}_{(i)}^\nu\partial^i_r\left(\dot{E}^\varphi-\{E^\varphi,H_T^{eff}[N,N_r]\}\right)\right] \nonumber \\
&+(\alpha_2^2-1)f_{(1)}^\nu+\frac{d\alpha_2}{dE^r}f_{(2)}^\nu.
\tag{\ref{DG1}}
\end{align*}
The coefficients are given by,

\begin{equation}
\bar{A}_{(0)}^\mu=\bar{A}_{(0)}n^\mu, \quad \bar{B}_{(0)}^\mu=\bar{B}_{(0)}n^\mu+l_{(1)}^\mu, \quad \bar{A}_{(1)}^\mu=\bar{A}_{(1)}n^\mu+l_{(2)}^\mu, \quad \bar{B}_{(1)}^\mu=\alpha_2B_{(1)}^\mu, \quad \bar{A}_{(2)}^\mu=\alpha_2A_{(2)}^\mu, \quad \bar{B}_{(2)}^\mu=0, \nonumber
\end{equation}
where the definitions of the $l^\mu$ vectors in (\ref{lmu}) are being utilized here too. In this case, additional terms responsible for the diffeomorphism invariance loss of this model appeared, namely the $f^\mu$ vectors in the second line of the past equation. These are, $$f_{(1)}^\mu=f_{(1)}n^\mu, \quad f_{(2)}^\mu=f_{(2)}n^\mu,$$ with

\begin{eqnarray}
f_{(1)}&=&\frac{1}{\gamma N\sqrt{\left|E^r\right|}(E^\varphi)^4}\left[NK_r\left((E^r)'^2E^\varphi-2E^r(E^r)'(E^\varphi)'\right)-(E^\varphi)^2\left(2N'K_\varphi'E^r+N(K_\varphi'E^r)'\right)\right. \nonumber \\
&&\hspace{2.8cm}\left.+E^rE^\varphi\left((NK_r(E^r)')'+NK_r(E^r)'+NK_\varphi'(E^\varphi)'\right)\right] \nonumber \\
f_{(2)}&=&\frac{\sqrt{\left|E^r\right|}(E^r)'}{\gamma(E^\varphi)^3}\left[K_r(E^r)'-K_\varphi'E^\varphi\right].
\end{eqnarray}

Using the emergent metric $g_{\mu\nu}^{(1)}$ of (\ref{ds2Esf2}), we expressed the corresponding divergence of the modified Einstein tensor in equation (\ref{DG2}), this is,

\begin{equation*}
\nabla_\mu\mathscr{G}^{\mu\nu}=\sum_{i=0}^2\left[\mathcal{A}_{(i)}^\nu\partial^i_r\left(\dot{E}^r-\{E^r,H_T^{eff}[N,N_r]\}\right)+\mathcal{B}_{(i)}^\nu\partial^i_r\left(\dot{E}^\varphi-\{E^\varphi,H_T^{eff}[N,N_r]\}\right)\right].
\tag{\ref{DG2}}
\end{equation*}
With coefficients given by,

\begin{equation}
\mathcal{A}_{(0)}^\mu=\frac{\bar{A}_{(0)}}{\alpha_2}n^\mu, \quad \mathcal{B}_{(0)}^\mu=\frac{\bar{B}_{(0)}}{\alpha_2}n^\mu+\alpha_2l_{(1)}^\mu, \quad \mathcal{A}_{(1)}^\mu=\frac{\bar{A}_{(1)}}{\alpha_2}n^\mu+\alpha_2l_{(2)}^\mu, \quad \mathcal{B}_{(1)}^\mu=B_{(1)}^\mu, \quad \mathcal{A}_{(2)}^\mu=A_{(2)}^\mu, \quad \bar{B}_{(2)}^\mu=0. \nonumber
\end{equation}
Note that these expressions also appear in the $\nabla_\mu\mathscr{G}^{\mu\nu}$ of (\ref{DG5}).

\subsection{C. Holonomy Corrections}

Again, we start by defining the quantities

\begin{eqnarray}
\widetilde{A}_{(0)}&=&\frac{1}{2\gamma N^2E^r(\bar{E}^\varphi)^3}\left[\gamma\bar{E}^\varphi\frac{\left(N'(E^r)'+2N''E^r\right)}{\sqrt{\left|\cos(\delta K_\varphi)\right|}}-\frac{2\gamma N'E^r(\bar{E}^\varphi)'}{\sqrt{\left|\cos(\delta K_\varphi)\right|}}+2\sqrt{\left|E^r\right|}(\bar{E}^\varphi)^2\left((N_rK_r)'-\dot{K}_r\right)\right. \nonumber \\
&&\hspace{2.5cm}\left.+\gamma\delta N'\bar{K}_\varphi'E^r\bar{E}^\varphi\frac{\tan(\delta K_\varphi)}{\left|\cos(\delta K_\varphi)\right|}\right] \nonumber \\
\widetilde{B}_{(0)}&=&\frac{1}{2\gamma N^2E^r(\bar{E}^\varphi)^4}\left[\gamma\bar{E}^\varphi\frac{\left[2\left(NE^r(E^r)'\right)'-N(E^r)'^2\right]}{\sqrt{\left|\cos(\delta K_\varphi)\right|}}-\frac{4\gamma NE^r(E^r)'(\bar{E}^\varphi)'}{\sqrt{\left|\cos(\delta K_\varphi)\right|}}+2\sqrt{\left|E^r\right|}(\bar{E}^\varphi)^3\left(N_rK_\varphi'-\dot{K}_\varphi\right)\right. \nonumber \\
&&\hspace{2.5cm}\left.+\gamma\delta N\bar{K}_\varphi'E^r(E^r)'\bar{E}^\varphi\frac{\tan(\delta K_\varphi)}{\left|\cos(\delta K_\varphi)\right|}\right], \nonumber \\
\widetilde{A}_{(1)}&=&\frac{1}{2N(\bar{E}^\varphi)^3E^r}\left[\frac{2E^r\left(\bar{E}^\varphi\right)'-\left(E^r\right)'\bar{E}^\varphi}{\sqrt{\left|\cos(\delta K_\varphi)\right|}}+\delta\bar{K}_\varphi'E^r\bar{E}^\varphi\frac{\tan(\delta K_\varphi)}{\left|\cos(\delta K_\varphi)\right|}\right],
\end{eqnarray}
which reduce to those of (\ref{ACl}) when $\delta=0$. In subsection IV.C., the emergent metric $g_{\mu\nu}^{(3)}$ of equation (\ref{ds2hol}) was used to calculate the divergence of the holonomy modified Einstein tensor. This was shown in equation (\ref{DG3}),

\begin{equation*}
\nabla_\mu\mathscr{G}^{\mu\nu}=\sum_{i=0}^2\left[\widetilde{A}_{(i)}^\nu\partial^i_r\left(\dot{E}^r-\{E^r,H_T^{eff}[N,N_r]\}\right)+\widetilde{B}_{(i)}^\nu\partial^i_r\left(\dot{\bar{E}}^\varphi-\{\bar{E}^\varphi,H_T^{eff}[N,N_r]\}\right)\right]+Fn^\nu.
\tag{\ref{DG3}}
\end{equation*}
The vectors can be written as,

\begin{align}
\widetilde{A}_{(0)}^\mu=&\widetilde{A}_{(0)}n^\mu, \quad \widetilde{B}_{(0)}^\mu=\widetilde{B}_{(0)}n^\mu+\widetilde{l}_{(1)}^\mu, \quad \widetilde{A}_{(1)}^\mu=\widetilde{A}_{(1)}n^\mu+\widetilde{l}_{(2)}^\mu, \nonumber \\
\widetilde{B}_{(1)}^\mu=&\frac{\left(E^r\right)'}{N\left(\bar{E}^\varphi\right)^3\sqrt{\left|\cos(\delta K_\varphi)\right|}}n^\mu, \quad \widetilde{A}_{(2)}^\mu=-\frac{1}{\left(\bar{E}^\varphi\right)^2\sqrt{\left|\cos(\delta K_\varphi)\right|}}n^\mu, \quad \widetilde{B}_{(2)}^\mu=0, \nonumber
\end{align}
where the notation $\widetilde{l}_{(1,2)}^\mu$ means that the variables $K_\varphi\rightarrow\bar{K}_\varphi$ and $E^\varphi\rightarrow\bar{E}^\varphi$ should be changed in the definitions of equations (\ref{lmu}).

Finally, when performing the same process but using the metric $g_{\mu\nu}^{(4)}$ of (\ref{ds2Esf4}), equation (\ref{DG4}) was obtained:

\begin{align*}
\nabla_\mu\mathscr{G}^{\mu\nu}=&\left(\left|\frac{\partial\gamma_2}{\partial K_\varphi}\right|\right)^{-3/2}\sum_{i=0}^2\left[A_{(i)}^\nu\partial^i_r\left(\dot{E}^r-\{E^r,H_T^{eff}[N,N_r]\}\right)+B_{(i)}^\nu\partial^i_r\left(\dot{E}^\varphi-\{E^\varphi,H_T^{eff}[N,N_r]\}\right)\right] \nonumber \\
&+K^\nu\mathcal{H}_{eff}.
\tag{\ref{DG4}}
\end{align*}
The vectors $A_{(i)}^\mu$ and $B_{(i)}^\mu$ were already defined in (\ref{ACl}). Here we give the expressions for $K^\mu$, namely, $$K^\mu=\frac{1}{4N\sqrt{\left|E^r\right|}E^\varphi}\left(\left|\frac{\partial\gamma_2}{\partial K_\varphi}\right|\right)^{-5/2}\left(\frac{\partial^2\gamma_2}{\partial K_\varphi^2}\right)\left(K_{(0)}n^\mu+K_{(1)}^\mu\right),$$ with 

\begin{equation}
K_{(0)}=N_rK_\varphi'-\dot{K}_\varphi, \quad K_{(1)}^\mu=\left[0,\frac{N^2K_\varphi'E^r}{\left(E^\varphi\right)^2}\frac{\partial\gamma_2}{\partial K_\varphi},0,0\right]. 
\end{equation} 
This concludes the complementary calculations of the main text. \\

\textbf{Acknowledgments.} JCDA acknowledges financial support from CONAHCyT postdoctoral fellowships. This work was also partially supported by CONAHCyT grant CBF-2023-2024-1937, as well as SNII-786529 (JCDA) and SNII-14585 (HAMT). \\

\section{References}

\bibliography{referencias}

\end{document}